\documentclass[12pt,a4paper]{article}
\usepackage[top=1in,bottom=1in,left=1in,right=1in,a4paper]{geometry}
\usepackage{epsfig,amsmath,amssymb,latexsym,graphicx,setspace,fullpage,amsthm,algorithm,sectsty}
\usepackage{array,rotating,multirow,enumitem,url,booktabs}
\newcolumntype{L}{>{\centering\arraybackslash}m{21mm}}
\newcolumntype{K}{>{\centering\arraybackslash}m{19mm}}

\usepackage{natbib}
\usepackage{float} 
\floatstyle{plain}
\newfloat{Algorithm}{thp}{lop}
\floatname{Algorithm}{Algorithm}

\newcommand{\ra}[1]{\renewcommand{\arraystretch}{#1}}
\usepackage[font={footnotesize}]{caption}
\sectionfont{\large}
\subsectionfont{\normalsize}

\begin{document}

\begin{center}
{\Large A stochastic variational framework for fitting and \\ diagnosing generalized linear mixed models}
\end{center}

\begin{center}
{\large Linda S. L. Tan and David J. Nott
\footnote{Linda S. L. Tan is research fellow (email statsll@nus.edu.sg) and David J. Nott is Associate Professor (email standj@nus.edu.sg), Department of Statistics and Applied Probability, National University of Singapore, Singapore 117546.}}
\end{center}

\begin{abstract}
In stochastic variational inference, the variational Bayes objective function is optimized using stochastic gradient approximation, where gradients computed on small random subsets of data are used to approximate the true gradient over the whole data set. This enables complex models to be fit to large data sets as data can be processed in mini-batches. In this article, we extend stochastic variational inference for conjugate-exponential models to nonconjugate models and present a stochastic nonconjugate variational message passing algorithm for fitting generalized linear mixed models that is scalable to large data sets. In addition, we show that diagnostics for prior-likelihood conflict, which are useful for Bayesian model criticism, can be obtained from nonconjugate variational message passing automatically, as an alternative to simulation-based Markov chain Monte Carlo methods. Finally, we demonstrate that for moderate-sized data sets, convergence can be accelerated by using the stochastic version of nonconjugate variational message passing in the initial stage of optimization before switching to the standard version. 
\end{abstract}


\section{Introduction}
Generalized linear mixed models (GLMMs) extend generalized linear models (GLMs) by introducing random effects to account for within-subject association and have wide applications. Estimation of GLMMs using maximum likelihood is, however, challenging as the integrals over random effects are intractable and have to be approximated using computationally intensive methods such as numerical quadrature or Markov chain Monte Carlo (MCMC). Various approximate methods for fitting GLMMs have been proposed, such as penalized quasi-likelihood \citep{Breslow1993}, Laplace approximation and its extensions \citep{Raudenbush2000}, Gaussian variational approximation \citep{Ormerod2012} and integrated nested Laplace approximations \citep{Fong2010}. Stochastic approximation has also been used in conjunction with MCMC \citep{Zhu2002} and the expectation maximization (EM) algorithm \citep{Jank2006} to fit GLMMs.

Recently, \citet{Tan2013} demonstrated how GLMMs can be fitted using variational Bayes \citep[VB,][]{Attias1999} via an algorithm called nonconjugate variational message passing \citep{Knowles2011}. A popular method of approximation, VB is deterministic and requires much less computation time than MCMC methods. In VB, the intractable true posterior is approximated by a factorized distribution, which is optimized to be close to the true posterior in terms of Kullback-Leibler divergence. Variational message passing \citep{Winn2005} is an algorithmic implementation of VB for conjugate-exponential models \citep{Ghahramani2001}. \cite{Knowles2011} extended variational message passing to nonconjugate models by assuming that the factors in VB belong to some exponential family.

The nonconjugate variational message passing algorithm for GLMMs \citep{Tan2013} has to update local variational parameters associated with every unit before re-estimating the global variational parameters at each iteration. This algorithm is inefficient for large data sets and is unsuitable for streaming data as it can never complete one iteration. To address these issues, \cite{Hoffman2013} proposed optimizing the VB objective function using stochastic gradient approximation \citep{Robbins1951}, where gradients computed on small random subsets of data are used to approximate the true gradient over the whole data set. This approach reduces the computational cost for large data sets significantly \citep{Bottou2005}. \cite{Hoffman2013} focused on developing stochastic variational inference for conjugate-exponential models.

In this article, we extend stochastic variational inference to nonconjugate models and develop a stochastic nonconjugate variational message passing algorithm for fitting GLMMs that is scalable to large data sets. A strong motivation for developing stochastic gradient optimization algorithms is their efficiency in terms of memory. As data are processed in mini-batches, analysis of data sets too large to fit into memory can still be contemplated. We focus on Poisson and logistic GLMMs, and applications in longitudinal data analysis. Our paper makes three contributions. First, we show how updates in nonconjugate variational message passing can be used in stochastic natural gradient optimization of the variational lower bound. Second, we show that variational message passing facilitates an automatic computation of diagnostics for prior-likelihood conflict (useful for Bayesian model criticism) and provides an attractive alternative to simulation-based MCMC methods. Third, we demonstrate that for moderate-sized data sets, convergence can be accelerated by using the stochastic version of nonconjugate variational message passing in the initial stage of optimization before switching to the standard version. 

Recently, there is increasing interest in developing VB algorithms capable of handling large data sets or streaming data \citep[e.g.][]{Luts2013, Broderick2013}. Stochastic optimization is an important tool in parameter estimation for large data sets \citep[e.g.][]{Bottou2008, Liang2013} and has been considered in the context of VB. For example, the online VB algorithms for latent Dirichlet allocation \citep{Hoffman2010} and the hierarchical Dirichlet process \citep{Wang2011} are based on stochastic natural gradient optimization of the VB objective function, with data processed one at a time or in mini-batches. \citet{Hoffman2013} generalized these methods to derive stochastic variational inference for conjugate-exponential family models. Stochastic approximation methods have also been considered by \citet{Ji2010}, \citet{Nott2012} and \citet{Paisley2012} for optimization of VB objective functions containing intractable integrals. \citet{Salimans2013} proposed a stochastic approximation algorithm that does not require analytic evaluation of integrals and allows fixed-form VB to be applied to any posterior available in closed form up to the proportionality constant. \cite{Nott2013} considers the approach of \cite{Salimans2013} for fitting GLMMs, and analyzes large data sets by combining variational approximations learned in parallel on smaller partitions. Random effects in each partition were treated as a single block. In this paper, we consider a different approach of fitting GLMMs to large data sets by using nonconjugate variational message passing within stochastic variational inference. Variational posteriors of random effects from different clusters are assumed to be independent and partial noncentering \citep{Tan2013} is used to improve posterior approximation. Global variational parameters are then updated using stochastic gradient approximation based on mini-batches of optimized local variational parameters. 

Model checking is an important part of statistical analyses. In the Bayesian approach, assumptions are made about the sampling model and prior, and prior-likelihood conflict arises when the observed data are very unlikely under the prior model. \citet{Evans2006} discuss how to assess whether there is prior-data conflict and \citet{Scheel2011} proposed a graphical diagnostic, the local critique plot, for identifying influential statistical modelling choices at the node level. See also \citet{Scheel2011} for a review of other methods in Bayesian model criticism. \citet{Marshall2007} proposed a diagnostic test for identifying divergent units in hierarchical models, based on measuring the conflict between the likelihood of a parameter and its predictive prior given the remaining data. A simulation-based approach was adopted and diagnostic tests were carried out using MCMC. We show that the approach of \citet{Marshall2007} can be approximated in the variational message passing framework.

Section 2 introduces some notation. Section 3 specifies the model and motivates partial noncentering for GLMMs. The stochastic nonconjugate variational message passing algorithm is developed in Section 4. Section 5 describes how variational message passing facilitates computation of prior-likelihood conflict diagnostics. Section 6 considers examples including real and simulated data and Section 7 concludes.

\section{Notation} \label{Notation}
We use $1_{d}$ to denote the $d \times 1$ column vector with all entries equal to 1 and $I_d$ to denote the $d \times d$ identity matrix. Scalar functions such as $\exp(\cdot)$ applied to vector arguments are evaluated element by element. We use $\odot$ to denote element by element multiplication of two vectors. If $a$ is a  $d \times 1$ vector, we use $\text{diag}(a)$ to denote the $d \times d$ diagonal matrix with diagonal entries given by $a$. On the other hand, if $A$ is a $d \times d$ square matrix, we use $\text{diag}(A)$ to denote the $d \times 1$ vector containing the diagonal entries of $A$.

\section{Generalized linear mixed models}\label{GLMM}
We consider one-parameter exponential family models which are specified as follows. Let $y_{ij}$ denote the $j$th response in cluster $i$, $i=1, \dots, n$, $j=1,\dots,n_i$. Conditional on a vector of length $r$ of random effects $u_i$, independently distributed as $N(0,D)$, $y_{ij}$ is independently distributed as
\begin{equation*}
y_{ij}|u_i \sim \exp\left\{y_{ij}\zeta_{ij}-b(\zeta_{ij})+c(y_{ij}) \right\},
\end{equation*}
where $\zeta_{ij}$ is the canonical parameter and $b(\cdot)$ and $c(\cdot)$ are functions specific to the exponential family. The link function $g$ relates the conditional mean of $y_{ij}$, $\mu_{ij}=E(y_{ij}|u_i)$ to the linear predictor $\eta_{ij}=x_{ij}^T \beta+z_{ij}^T u_i$ as $g(\mu_{ij})=\eta_{ij}$. Here, $x_{ij}$ and $z_{ij}$ are $p \times 1$ and $r \times 1$ vectors of covariates and $\beta$ is a $p \times 1$ vector of unknown fixed regression parameters. We consider responses from the Bernoulli and Poisson families. If $y_{ij}\sim \text{Bernoulli}(\mu_{ij})$, then $b(x)=\log\{1+\exp(x)\}$, $c(x)=0$ and $\text{logit}(\mu_{ij})=\eta_{ij}$. For Poisson responses, we allow for an offset $\log E_{ij}$. If $y_{ij}\sim \text{Poisson}(\mu_{ij})$, then $b(x) = \exp(x)$, $c(x)=-\log(x!)$ and $\log \mu_{ij} = \log E_{ij} +\eta_{ij}$. For the $i$th cluster, let 
\begin{equation*}
y_i = \begin{bmatrix} y_{i1}\\ \vdots \\ y_{in_i} \end{bmatrix}, \;
\eta_i = \begin{bmatrix} \eta_{i1} \\ \dots \\ \eta_{in_i} \end{bmatrix}, \;
X_i = \begin{bmatrix} x_{i1}^T \\ \vdots \\ x_{in_i}^T \end{bmatrix}, \;
Z_i = \begin{bmatrix} z_{i1}^T \\ \vdots \\ z_{in_i}^T \end{bmatrix}\;\; \text{and}\;\;
E_i = \begin{bmatrix} E_{i1}\\ \vdots \\ E_{in_i} \end{bmatrix}, \;
\end{equation*}
We assume that the first column of $Z_i$ is $1_{n_i}$ if $Z_i$ is not a zero matrix and that the columns of $Z_i$ are a subset of the columns of $X_i$. 

For Bayesian inference, we specify a diffuse prior $N(0,\Sigma_\beta)$ on $\beta$ where $\Sigma_\beta$ is large and an independent inverse-Wishart prior, $IW(\nu,S)$ on $D$. We use the default conjugate prior proposed in \citet{Kass2006}, which is based on a prior guess for $D$ determined from first-stage data variability. For this default prior, $\nu=r$ and $S=r\hat{R}$ where 
\begin{equation}\label{Kassprior}
\hat{R}=c\left(\frac{1}{n}\sum_{i=1}^{n}Z_i^T M_i(\hat{\beta})Z_i\right)^{-1}.
\end{equation}
Here, $M_i(\hat{\beta})$ denotes the $n_i \times n_i$ diagonal GLM weight matrix with diagonal elements $[v(\hat{\mu}_{ij})\, g'(\hat{\mu}_{ij})^2]^{-1}$, where $v(\cdot)$ is the variance function and $g(\cdot)$ is the link function. We let $\hat{\mu}_{ij}=g^{-1}(x_{ij}^T\hat{\beta} + z_{ij}^T\hat{u}_i)$ where $\hat{u}_i$ is set as $0$ for all $i$ and $\hat{\beta}$ is an estimate of the regression coefficients from the GLM obtained by pooling all data and setting $u_i=0$ for all $i$. The constant $c$ is an inflation factor representing the amount in which within-cluster variability can be increased. We use $c=1$ in all examples. Some heuristic justifications for $\hat{R}$ is given in \citet{Kass2006}. A similar prior was used in \cite{Overstall2010}. Alternatively, one may consider marginally noninformative priors for covariance matrices \citep{Huang2013}. Methods in this paper can be extended to these priors easily.

\subsection{A partially noncentered parametrization for the GLMM}\label{Sec_PNCP}

Reparametrization techniques such as centering, noncentering and partial noncentering have been used in hierarchical models to boost efficiency in MCMC and EM algorithms \citep[e.g.][]{Gelfand1995, Gelfand1996, Papaspiliopoulos2003, Papaspiliopoulos2007}. Recently, \cite{Tan2013} introduced a partially noncentered parametrization for GLMMs and studied its performance in the context of VB. We introduce the idea of partial noncentering by considering the following linear mixed model \citep[see also][]{Tan2013}. Suppose
\begin{equation} \label {LMM}
y_i=X_i \beta+ Z_i u_i+\epsilon_i \;\; \text{where}\;\; \epsilon_i \sim N(0,\sigma^2) \;\;\text{for}\;\; i=1, \dots,n,
\end{equation} 
and $y_i$, $X_i$, $Z_i$, $u_i$ and $\beta$ are as defined previously. Let us specify a constant prior on $\beta$ and assume that $\sigma^2$ and $D$ are known. Suppose $X_i=Z_i$. In this case, we can introduce $\alpha_i=\beta+u_i$ so that $\alpha_i \sim N(\beta,D)$ is ``centered'' about $\beta$. We can also obtain a partially noncentered parametrization by introducing $\tilde{\alpha}_i=\alpha_i-W_i \beta$, where $W_i$ is an $r \times r$ tuning matrix to be specified. The proportion of $\beta$ subtracted from $\alpha_i$ is allowed to vary with $i$ as each $y_i$ carries different amount of information about the underlying $\alpha_i $. The centered ($W_i=0$) and noncentered ($W_i=I_r$) parametrizations are special cases of the partially noncentered parametrization. Rewriting \eqref{LMM} as
\begin{equation*} 
y_i=Z_iW_i \beta+ Z_i \tilde{\alpha}_i  +\epsilon_i,
\end{equation*} 
we can apply VB to the reparametrized model and assume that $q(\beta,\tilde{\alpha}_1,\dots,\tilde{\alpha}_n)=q(\beta)\prod_{i=1}^n q(\tilde{\alpha}_i)$. \cite{Tan2013} showed that the resulting VB algorithm converges in one iteration when
\begin{equation} \label{tuning}
W_i=(Z_i^T Q_i Z_i+D^{-1})^{-1}D^{-1},
\end{equation}
where $Q_i = \frac{1}{ \sigma^2} I_r$. This result implies that partial noncentering can yield more rapid convergence than centering or noncentering. More importantly, the true posteriors are recovered in \eqref{tuning} but not in the centered or noncentered parametrizations. Even though assumption of a factorized posterior in VB tends to result in underestimation of posterior variance, partial noncentering was (in this case) able to capture dependence between fixed and random effects via tuning parameters $W_i$ so that the true posterior can be recovered.

The above result is particularly useful in the context of stochastic variational inference for GLMMs. To implement stochastic variational inference, we need to assume that variational posteriors of random effects associated with each unit are independent of each other and of the global variables $\beta$ and $D$. However, correlation between fixed and random effects is often strong and partial noncentering allows some of this dependence to be captured via the tuning matrices $W_i$. This leads to more accurate posterior approximations of the fixed and random effects. In particular, estimation of the posterior variance of fixed effects which can be centered is improved greatly. Partial noncentering can also give more rapid convergence than centering or noncentering. This is desirable in the analysis of large data sets and is particularly useful when the convergence of one of the centered or noncentered parametrizations is especially slow. We emphasize that it is not easy to tell beforehand which of centering or noncentering will perform better, and partial noncentering automatically chooses a parametrization close to optimal.

We adopt the partially noncentered parametrization introduced by \cite{Tan2013} for the GLMM, which is explained below. First, we partition $X_i$ as $[\begin{matrix} Z_i & X_{si} & X_{gi} \end{matrix}]$ and $\beta$ as $[\beta_z^T, \beta_s^T, \beta_g^T]^T$ accordingly, where $X_{si}$ is a $n_i \times s$ matrix consisting of ``subject specific'' covariates and $X_{gi}$ is a $n_i \times g$ matrix consisting of ``general'' covariates (i.e. not subject specific). All the rows of $X_{si}$ are thus the same and equal to say $x_{si}^T$. We have
\begin{align*}
\eta_i &= Z_i (\beta_z+u_i) + 1_{n_i}x_{si}^T \beta_s +  X_{gi} \beta_g  \\
      &= Z_i ( C_i \beta_c+ u_i) +  X_{gi}\beta_g, \;\; \text{where}\;\; 
C_i =\left[\begin{matrix}  I_r & \begin{matrix} x_{si}^T \\ 0_{(r-1)\times s}\end{matrix}  \end{matrix}\right] \text{and}\; \beta_c=\begin{bmatrix} \beta_z \\ \beta_s \end{bmatrix}.
\end{align*}
Note that $C_i$ is an $r \times (r+s)$ matrix. We introduce 
\begin{equation*}
\alpha_i = C_i \beta_c+ u_i \;\; \text{and} \;\; \tilde{\alpha}_i=\alpha_i-W_i C_i \beta_c,
\end{equation*}
where $W_i$ is an $r \times r$ tuning matrix. $W_i=0$ corresponds to the centered and $W_i=I_r$ to the noncentered parametrization. Letting $\tilde{W}_i=[ \begin{matrix} (I_r-W_i)C_i & 0_{r\times g} \end{matrix} ]$ be an $r \times p$ matrix, $\tilde{\alpha}_i \sim N(\tilde{W}_i \beta,D)$. The partially noncentered parametrization is thus
\begin{equation*}
\eta_i  = V_i \beta +Z_i \tilde{\alpha}_i,
\end{equation*}
where $V_i = [\begin{matrix} Z_i W_iC_i & X_{gi} \end{matrix}]$ is a $n_i \times p$ matrix. Following \citet{Tan2013}, $W_i$ can be specified as in \eqref{tuning} with $Q_i = \text{diag}\left(\frac{\exp(\eta_i)}{\{1+\exp(\eta_i)\}^2}\right)$ for logistic GLMMs and $Q_i =\text{diag}(E_i\odot \eta_i)\approx \text{diag}(y_i)$ for Poisson GLMMs. 

Let $y=[y_1^T,\dots,y_n^T]^T$ and $\tilde{\alpha}=[\tilde{\alpha}_1^T,\dots,\tilde{\alpha}_n^T]^T$. The set of unknown parameters in the GLMM is $\theta=\{\beta, D, \tilde{\alpha}\}$ and
\begin{equation}\label{prob_distn}
p(y,\theta) =\left\{ \prod_{i=1}^n p(y_i|\beta,\tilde{\alpha}_i)p(\tilde{\alpha}_i|\beta,D) \right\} p(\beta|\Sigma_\beta) p(D|\nu,S).
\end{equation}
The fixed effects $\beta$ and random effects covariance $D$ can be regarded as ``global'' variables which are common across clusters, while the partially noncentered random effects $\tilde{\alpha}_i$ can be thought of as ``local'' variables associated only with the individual units.

\section{Stochastic variational inference for generalized linear mixed models}
In this section, we derive and present the stochastic nonconjugate variational message passing algorithm for fitting GLMMs, which is scalable to large data sets. We start with a brief introduction to variational approximation methods \citep[see, e.g.][]{Ormerod2010} and review of nonconjugate variational message passing \citep{Knowles2011}. 

In variational approximation, the true posterior $p(\theta|y)$ is approximated by a more tractable distribution $q(\theta|\lambda)$, where $\lambda$ denotes the set of parameters of $q$. We attempt to make $q(\theta|\lambda)$ a good approximation to $p(\theta|y)$ by minimizing the Kullback-Leibler divergence between $q(\theta|\lambda)$ and $p(\theta|y)$. This is given by
\begin{equation*}
\int q(\theta|\lambda)\log \frac {q(\theta|\lambda)}{p(\theta|y)} \,d\theta 
= \int q(\theta|\lambda)\log \frac{q(\theta|\lambda)}{p(y,\theta)} \,d\theta + \log p(y),
\end{equation*} 
where $p(y)=\int p(y,\theta) \,d\theta$ is the marginal likelihood. As the Kullback-Leibler divergence is nonnegative, we have 
\begin{equation*}
\begin{aligned}
\log p(y) &\geq \int q(\theta|\lambda)\log \frac {p(y,\theta)}{q(\theta|\lambda)} \,d\theta \\
          & = E_q\{\log p(y,\theta)\}-E_q\{\log q(\theta|\lambda)\} = \mathcal{L},
\end{aligned}
\end{equation*}
where $E_q$ denotes expectation with respect to $q(\theta|\lambda)$ and $\mathcal{L}$ is a lower bound  on the log marginal likelihood. Maximization of $\mathcal{L}$ is thus equivalent to minimization of the Kullback-Leibler divergence between $q(\theta|\lambda)$ and $p(\theta|y)$. In some cases, $\mathcal{L}$ is used as an approximation to the log marginal likelihood for performing model selection. See \cite{Tan2013} for an illustration of how $\mathcal{L}$ can be used for model selection in GLMMs.

\subsection{Nonconjugate variational message passing} 
In VB, $q(\theta|\lambda)$ is assumed to factorize into $\prod_{l=1}^m q_l(\theta_l|\lambda_l)$ for some partition $\{\theta_1,\dots,\theta_m\}$ of $\theta$ and $\lambda_l$ denotes variational parameters associated with each factor. Optimization of $\mathcal{L}$ with respect to $q_1, \dots, q_m$ leads to optimal densities satisfying 
\begin{equation}\label{VB_opt}
q_l(\theta_l) \propto \exp [E_{-\theta_l} \{\log p(y,\theta)\}],
\end{equation} 
where $E_{-\theta_l}$ denotes expectation with respect to $\prod_{j \neq l} q_j(\theta_j|\lambda_j)$. When conjugate priors are used, the optimal densities have the same form as the priors and it suffices to update the parameters of $q_l$. However, for non-conjugate priors, the optimal densities may not belong to recognizable density families. To address this issue, \cite{Knowles2011} imposed a further restriction that each $q_l$ must belong to some exponential family. Let 
\begin{equation*}
q_l(\theta_l|\lambda_l)=\exp\{\lambda_l^T t_l(\theta_l)-h_l(\lambda_l)\},
\end{equation*}
where $\lambda_l$ is the vector of natural parameters and $t_l(\cdot)$ are the sufficient statistics. Updates in nonconjugate variational message passing can be derived by maximizing $\mathcal{L}$ with respect to each $\lambda_l$ and setting $\nabla_{\lambda_l}\mathcal{L}=0$. Let $\mathcal{V}_l(\lambda_l)$ denote the covariance matrix of $t_l(\theta_l)$. It can be shown that
\begin{equation}\label{reg_grad}
\nabla_{\lambda_l}\mathcal{L} =\nabla_{\lambda_l} E_q\{\log p(y,\theta)\}-\mathcal{V}_l(\lambda_l)\lambda_l.
\end{equation}
Updates in nonconjugate variational message passing are thus given by 
\begin{equation} \label{NCVMP_update}
\lambda_l \leftarrow \mathcal{V}(\lambda_l)^{-1} \nabla_{\lambda_l} E_q\{\log p(y,\theta)\}
\end{equation}
for $l=1,\dots,m$. As nonconjugate variational message passing is a type of fixed-point iterations algorithm, the lower bound is not guaranteed to increase after each update. Sometimes, convergence issues may be encountered which may require damping to fix \citep[see][]{Knowles2011}. For conjugate factors, the update in \eqref{NCVMP_update} can be simplified and details are given in Appendix A.

The nonconjugate variational message passing algorithm for GLMMs \citep{Tan2013} considers a variational approximation of the form 
\begin{equation}\label{VB_GLMM}
q(\theta|\lambda)=q(\beta|\lambda_\beta)q(D|\lambda_D)\prod_{i=1}^n  q(\tilde{\alpha}_i|\lambda_{\tilde{\alpha}_i}),
\end{equation}
where $q(\beta|\lambda_\beta)$ is $N(\mu_{q(\beta)},\Sigma_{q(\beta)})$, $q(D|\lambda_D)$ is $IW(\nu_{q(D)},S_{q(D)})$, $q(\tilde{\alpha}_i|\lambda_{\tilde{\alpha}_i})$ is $N(\mu_{q(\tilde{\alpha}_i)},\Sigma_{q(\tilde{\alpha}_i)})$ and $\lambda_\beta$, $\lambda_D$, $\lambda_{\tilde{\alpha}_i}$ are the respective natural parameter vectors. For Bernoulli or Poisson responses, $p(y_i|\beta,\tilde{\alpha}_i)$ is nonconjugate with respect to the priors over $\beta$ and $\tilde{\alpha}_i$. Applying nonconjugate variational message passing and approximating the posteriors of $\beta$ and $\tilde{\alpha}_i$ by Gaussian distributions, parameter updates for $q(\beta)$ and $q(\tilde{\alpha}_i)$ can be derived using \eqref{NCVMP_update}. The variational posterior for $D$ is optimal under \eqref{VB_GLMM} and parameter updates can be derived using \eqref{VB_opt}. The main steps are given in Algorithm \ref{Alg 1} below.
\begin{Algorithm}
\centering
\parbox{0cm}{
\hrule\begin{tabbing}
\\[-8mm]
\=Initialize $\mu_{q(\beta)}$, $\Sigma_{q(\beta)}$, $\nu_{q(D)}$, $S_{q(D)}$, $\mu_{q(\tilde{\alpha}_i)}$, $\Sigma_{q(\tilde{\alpha}_i)}$ and tuning parameters $W_i$ for $i=1,\dots,n$. \\
\> Cycle:\= \\
\>\>1.  Update local variational parameters $\mu_{q(\tilde{\alpha}_i)}$ and $\Sigma_{q(\tilde{\alpha}_i)}$ for each $i=1,\dots,n$. \\ [1mm]
\>\>2.  Update global variational parameters $\mu_{q(\beta)}$, $\Sigma_{q(\beta)}$, $\nu_{q(D)}$ and $S_{q(D)}$. \\ [1mm]
\> until the lower bound converges.
\\[-9mm]
\end{tabbing}\hrule}
\caption{\label{Alg 1} Nonconjugate variational message passing for GLMMs.}
\end{Algorithm}

Algorithm \ref{Alg 1} iterates repeatedly between updating local variational parameters for each unit $i=1,\dots,n$, and re-estimating the global variational parameters. This procedure is inefficient for large data sets and impossible to accomplish for streaming data or data sets too massive to fit into memory. Using ideas in stochastic variational inference \citep{Hoffman2013}, we develop a stochastic nonconjugate variational message passing algorithm for fitting GLMMs that is more efficient at handling large data.

\subsection{Natural gradient of the variational lower bound} \label{Sec_nat_grad}
In stochastic variational inference, the global variational parameters are optimized using stochastic gradient ascent. Updates of the form
\begin{equation*}
\lambda^{(t+1)} = \lambda^{(t)} + a_t \, \nabla_{\lambda} \mathcal{L} (\lambda^{(t)})
\end{equation*}
are considered, where $a_t$ denotes a small step taken in the direction of steepest ascent at the $t$th iteration. Under the Euclidean metric, the direction of steepest ascent is given by the regular gradient ${\nabla}_{\lambda} \mathcal{L} (\lambda^{(t)})$. In stochastic gradient ascent, a noisy estimate of ${\nabla}_{\lambda} \mathcal{L} (\lambda^{(t)})$ is used in its place. \cite{Hoffman2013} propose using natural gradients instead of regular gradients in this optimization step. Their motivation is that the Euclidean distance between two parameter settings $\lambda$ and $\lambda'$ is often a poor measure of how dissimilar two distributions $q(\theta|\lambda)$ and $q(\theta|\lambda')$ are. A more intuitive measure of dissimilarity between two probability distributions is given by the symmetrized Kullback-Leibler divergence, which is invariant to parameter transformations. Under this measure, \cite{Hoffman2013} showed that the direction of steepest ascent is given by the natural gradient \citep{Amari1998}. Previously, \citet{Honkela2008} also showed that replacing regular gradients with natural gradients in the conjugate gradient algorithm can speed up variational learning. 

 Following \cite{Hoffman2013}, we use natural gradients instead of regular gradients in the stochastic approximation. To obtain the natural gradient of $\mathcal{L}$ with respect to $\lambda_l$, we premultiply $\nabla_{\lambda_l} \mathcal{L}$ with the inverse of the Fisher information matrix of $q_l(\theta_l|\lambda_l)$ \citep[see, e.g.][]{Amari1998}. In nonconjugate variational message passing, the Fisher information matrix is given by
\begin{align*}
E_q\left[ \left\{ \nabla_{\lambda_l} \log q_l(\theta_l|\lambda_l) \right\} \left\{ \nabla_{\lambda_l} \log q_l(\theta_l|\lambda_l) \right\} ^T\right]
&= E_q\left[ \left\{  t_l(\theta_l)-\nabla_{\lambda_l} h_l(\lambda_l) \right\} \left\{ t_l(\theta_l)-\nabla_{\lambda_l}h_l(\lambda_l) \right\} ^T\right]  \\
&=\mathcal{V}_l(\lambda_l).
\end{align*}
From \eqref{reg_grad}, the natural gradient denoted by $\widetilde{\nabla}_{\lambda_l}\mathcal{L}$ is thus given by 
\begin{equation}\label{nat_grad}
\widetilde{\nabla}_{\lambda_l}\mathcal{L} = \mathcal{V}_l(\lambda_l)^{-1} \nabla_{\lambda_l} E_q\{\log p(y,\theta)\} - \lambda_l.
\end{equation}

\subsection{Stochastic variational inference} 
In this section, we review the key ideas in stochastic variational inference \citep{Hoffman2013} and discuss how they can be extended to nonconjugate models via nonconjugate variational message passing. The following steps are carried out in each iteration of stochastic variational inference until convergence is reached.
\begin{enumerate}
\item Randomly select a mini-batch $B$ of of $|B|\geq1$ units from the whole data set.
\item Optimize local variational parameters of units in mini-batch $B$ (as a function of the global variational parameters at their current setting).
\item Update global variational parameters using stochastic natural gradient ascent. Noisy gradients are computed based on optimized local variational parameters of units in mini-batch $B$.
\end{enumerate}

The main difficulty in extending stochastic variational inference to nonconjugate models lies in step 2. For conjugate models, the local variational parameters can be optimized as a function of the global variational parameters in a single update [see \eqref{VMP_update}] but the same is not true for nonconjugate models. In nonconjugate variational message passing, the update equation for the local variational parameters is recursive (they depend on the current setting of the local variational parameters) and has to be applied repeatedly until convergence is reached [see \eqref{NCVMP_update}]. This incurs a higher computational cost. We have tried performing the update for local variational parameters only once but this further slowed down convergence of the global variational parameters. We have also tried using a loose criterion for assessing convergence. This approach yielded much better results. Choosing a good initialization is also important as convergence problems can be encountered in recursive updates if the starting point is poor. 

The other main difference is that for conjugate models, the update equations and natural gradients are easier to compute as the Fisher information matrix $\mathcal{V}_i(\lambda_i)$ does not have to be evaluated [see \eqref{VMP_update} and \eqref{VMP_nat_grad}]. Fortunately, nonconjugate variational message passing updates can be simplified considerably when the variational posteriors are multivariate Gaussian \citep[see][]{Wand2013} and the Fisher information matrix does not have to be computed explicitly as well. 

The extension of stochastic variational inference to nonconjugate models greatly widens the scope of models to which stochastic variational inference can be applied. We think that nonconjugate variational message passing is an important tool in facilitating this extension as it allows for efficient closed-form updates in some cases (e.g. Poisson GLMMs) and there is a lot of flexibility in the evaluation of expectations (using bounds or quadrature). While convergence issues remain in fixed-point iterations algorithms, these can usually be mitigated by good initializations. We later show that nonconjugate variational message passing, like VB, is a type of natural gradient method \citep[see][]{Sato2001}. With this interpretation, some convergence issues might be resolved by taking adaptive steps in the direction of the natural gradient.

Let $\lambda_{\text{global}}$ and $\lambda_{\text{local}}$ denote the global and local variational parameters respectively. The lower bound $\mathcal{L}$ is a function of $\lambda = (\lambda_{\text{global}},\lambda_{\text{local}} )$, i.e. $\mathcal{L}=\mathcal{L}(\lambda)=\mathcal{L}(\lambda_{\text{global}},\lambda_{\text{local}} )$. \cite{Hoffman2013} showed that to find a setting of $\lambda_{\text{global}}$ that maximizes $\mathcal{L}$ using stochastic natural gradient ascent, we can first optimize $\lambda_{\text{local}}$ as a function of $\lambda_{\text{global}}$ so that $\lambda_{\text{local}} = k(\lambda_{\text{global}})$ for some function $k$. In nonconjugate variational message passing, this is done by computing the update in \eqref{NCVMP_update} repeatedly until convergence, starting with some current setting of $\lambda_{\text{local}}$ and keeping $\lambda_{\text{global}}$ fixed. This implies that $\nabla_k \mathcal{L}(\lambda_{\text{global}},k(\lambda_{\text{global}}))=0$ since $k(\lambda_{\text{global}})$ is a local optimum of the local variational parameters. The current value of the lower bound is $ \mathcal{L} (\lambda_{\text{global}},k(\lambda_{\text{global}}))$ which is a function of $\lambda_{\text{global}}$ only. Let us define $\mathcal{L}(\lambda_{\text{global}}) = \mathcal{L} (\lambda_{\text{global}},k(\lambda_{\text{global}})) $. To optimize $\mathcal{L}(\lambda_{\text{global}})$ with respect to $\lambda_{\text{global}}$, we have
\begin{align*}
\nabla\negthinspace_{\lambda_{\text{global}}} \mathcal{L} (\lambda_{\text{global}}) & =\nabla \negthinspace _{\lambda_{\text{global}}} \mathcal{L} (\lambda_{\text{global}},k(\lambda_{\text{global}})) + \{ \nabla\negthinspace_{\lambda_{\text{global}}} k(\lambda_{\text{global}}) \}^T \nabla\negthinspace_k \mathcal{L}(\lambda_{\text{global}},k(\lambda_{\text{global}}))  \\
& =\nabla\negthinspace_{\lambda_{\text{global}}} \mathcal{L} (\lambda_{\text{global}},k(\lambda_{\text{global}})).
\end{align*}
Therefore, $\nabla_{\lambda_{\text{global}}} \mathcal{L} (\lambda_{\text{global}})$ can be computed by finding the optimized local variational parameters $k(\lambda_{\text{global}})$ and then computing the gradient of $\mathcal{L} (\lambda_{\text{global}},k(\lambda_{\text{global}})) $ with respect to $\lambda_{\text{global}}$ by keeping $k(\lambda_{\text{global}})$ fixed. The corresponding natural gradient can be obtained as discussed in Section \ref{Sec_nat_grad}.

In stochastic variational inference, noisy estimates of the natural gradients are used in stochastic optimization of the global variational parameters. The idea is to approximate true gradients over the whole data with gradients computed on mini-batches of data. For large data sets, this can lead to significant reductions in computation time. For the GLMM, $\lambda_{\text{global}}=(\lambda_\beta, \lambda_D)$ and $\lambda_{\text{local}}=(\lambda_{\tilde{\alpha}_1}, \dots, \lambda_{\tilde{\alpha}_n})$. As $\beta$ and $D$ are independent  in the variational posterior, stochastic gradient ascent for $\lambda_\beta$ and $\lambda_D$ can be done separately. From \eqref{prob_distn} and \eqref{nat_grad}, the natural gradient of $\mathcal{L}$ with respect to $\lambda_\beta$, $\widetilde{\nabla}_{\lambda_\beta}\mathcal{L}$ is given by
\begin{equation}\label{nat_grad_true}
\mathcal{V}_\beta(\lambda_\beta)^{-1} \nabla_{\lambda_\beta} \bigg\{\sum_{i=1}^n E_q\{ \log p(y_i|\beta,\tilde{\alpha}_i) + \log p(\tilde{\alpha}_i|\beta,D)\}|_{\lambda_{{\tilde{\alpha}_i}}=  \lambda_{{\tilde{\alpha}_i}}^{\text{opt}}} + E_q\{\log p(\beta|\Sigma_\beta)\} \bigg\} - \lambda_\beta,
\end{equation}
where $\lambda_{{\tilde{\alpha}_i}}^{\text{opt}}$ denotes $\lambda_{{\tilde{\alpha}_i}}$ optimized as a function of the global variational parameters. If $B$ is a mini-batch of $|B|$ units randomly sampled from the whole data set (with or without replacement), then an unbiased estimate of $\widetilde{\nabla}_{\lambda_\beta}\mathcal{L} $ is $\hat{\lambda}_\beta  - \lambda_\beta$, where
\begin{equation*}
\hat{\lambda}_\beta = \mathcal{V}_\beta(\lambda_\beta)^{-1} \nabla_{\lambda_\beta} \bigg\{ \frac{n}{|B|} \sum_{i \in B} E_q\{ \log p(y_i|\beta,\tilde{\alpha}_i) + \log p(\tilde{\alpha}_i|\beta,D)\} |_{\lambda_{{\tilde{\alpha}_i}}=  \lambda_{{\tilde{\alpha}_i}}^{\text{opt}}} + E_q\{\log p(\beta|\Sigma_\beta)\}  \bigg\}.
\end{equation*}
Note that each of the $n$ units in the whole data set has a probability $\frac{|B|}{n}$ of being selected and hence the expectation of $\hat{\lambda}_\beta  - \lambda_\beta $ is equal to \eqref{nat_grad_true} \citep[][pg. 18 -- 19]{Hoffman2013}. Similarly, an unbiased estimate of $\widetilde{\nabla}_{\lambda_D}\mathcal{L} $ is $\hat{\lambda}_D - \lambda_D$, where
\begin{equation*} 
\hat{\lambda}_D = \mathcal{V}_D(\lambda_D)^{-1} \nabla_{\lambda_D} \bigg\{  \frac{n}{|B|} \sum_{i \in B} E_q\{\log p(\tilde{\alpha}_i|\beta,D)\} |_{\lambda_{{\tilde{\alpha}_i}}=  \lambda_{{\tilde{\alpha}_i}}^{\text{opt}}}  + E_q\{\log p(D|\nu,B)\}  \bigg\}.
\end{equation*}
When $B$ is the whole data set, $\hat{\lambda}_\beta$ and $\hat{\lambda}_D$ are respectively the updates of $\lambda_\beta$ and $\lambda_D$ in nonconjugate variational message passing.

With these unbiased estimates of the natural gradients, $\lambda_\beta$ and $\lambda_D$ can be updated using stochastic gradient approximation \citep{Robbins1951}. At the $t$th iteration,
\begin{equation} \label{SAupdates}
\lambda_\beta^{(t+1)}=\lambda_\beta^{(t)}+a_t\;(\hat{\lambda}_\beta  - \lambda_\beta^{(t)})  \quad\text{and}\quad \lambda_D^{(t+1)}=\lambda_D^{(t)}+a_t \; (\hat{\lambda}_D - \lambda_D^{(t)}), 
\end{equation}
where $\hat{\lambda}_\beta$ and $\hat{\lambda}_D$ are evaluated using the current settings of $\lambda_\beta$ and $\lambda_D$. Under certain regularity conditions \citep[see][]{Spall2003}, the iterates will converge to a local maximum of the lower bound. The gain sequence $a_t$, $t\geq 0$ should satisfy
\begin{gather*}
a_t\rightarrow 0,\;\;\;\sum_{t=0}^\infty a_t=\infty,\;\;\;\text{and}\;\;\;\sum_{t=0}^\infty a_t^2<\infty.
\end{gather*}
The condition $(a_t\rightarrow 0, \sum_{t=0}^\infty a_t^2<\infty)$ ensures that the step size goes to zero sufficiently fast so that iterates will converge while $(\sum_{t=0}^\infty a_t=\infty)$ ensures that the rate at which step sizes approach zero is slow enough to avoid false convergence. \cite{Spall2003} recommends 
\begin{equation} \label{Spallstepsize}
a_t = \frac{a}{(t+A)^\alpha},
\end{equation}
where $0.5<\alpha\leq 1$, $A \geq 0$ is a stability constant that helps to avoid unstable behaviour in the early iterations and $a$ keeps step sizes nonnegligible in later iterations. Note that updates in \eqref{SAupdates} can be rewritten as
\begin{equation}\label{previousiterate}
\lambda_\beta^{(t+1)}=(1-a_t)\, \lambda_\beta^{(t)}+a_t \, \hat{\lambda}_\beta \quad\text{and}\quad  \lambda_D^{(t+1)}=(1-a_t)\,\lambda_D^{(t)}+a_t \,\hat{\lambda}_D.
\end{equation} 
The $t$th iterate is thus a weighted average of the previous iterate and the nonconjugate variational message passing update estimated using mini-batch $B$. When $a_t=1$ and $B$ is the whole data set,  $\lambda_\beta^{(t)}$ is precisely the update in nonconjugate variational message passing. This shows that nonconjugate variational message passing is a type of natural gradient method with step size 1 and other schedules are equivalent to damping.

\subsection{Stochastic nonconjugate variational message passing algorithm}\label{SNCVMP}
The stochastic nonconjugate variational message passing algorithm for fitting Poisson and logistic GLMMs is presented in Algorithm \ref{Alg 2}. Derivation of updates and definitions of $F_i$ and $g_i$ (appearing in Algorithm \ref{Alg 2} and which differ according to whether logistic or Poisson GLMMs are fitted) are given in Appendix B. Algorithm \ref{Alg 2} reduces to Algorithm \ref{Alg 1} when mini-batch $B$ is the entire data set, $a_t=1$, and updates for local variational parameters are performed only once. 
\begin{Algorithm}[t]
\centering
\parbox{0cm}{
\hrule\begin{tabbing}
\\[-8mm]
\=Initialize $\mu_{q(\beta)}$, $\Sigma_{q(\beta)}$, $S_{q(D)}$, $\mu_{q(\tilde{\alpha}_i)}$, $\Sigma_{q(\tilde{\alpha}_i)}$ and tuning parameters $W_i$ for $i=1,\dots,n$. \\ [1mm]
\> Set $\nu_{q(D)}=\nu+n$.  \\ [1mm]
\> For \= $t$ =  0, 1, 2, $\ldots$,  \\ [1mm]
\>\>1. \= Draw a mini-batch $B$ of $|B| \geq 1$ units from the whole data set at random and \\
\> \> \> without replacement. \\ [1mm]
\>\>2. Update local variational parameters $\mu_{q(\tilde{\alpha}_i)}$ and $\Sigma_{q(\tilde{\alpha}_i)}$ for $i \in B$ repeatedly \\
\>\>\>   using: \=  \\
\>\>\>\>   $\Sigma_{q(\tilde{\alpha}_i)} \leftarrow \big(\nu_{q(D)} S_{q(D)}^{-1}+ Z_i^T F_i Z_i \big)^{-1}$  \\ [1mm]
\>\>\>\>   $\mu_{q(\tilde{\alpha}_i)} \leftarrow \mu_{q(\tilde{\alpha}_i)} + \Sigma_{q(\tilde{\alpha}_i)} \big\{Z_i^T (y_i-g_i)-\nu_{q(D)}S_{q(D)}^{-1}(\mu_{q(\tilde{\alpha}_i)}-\tilde{W}_i\mu_{q(\beta)}) \big\} $  \\ [1mm]
\>\>\>   until convergence is reached.   \\ [1mm]
\>\>3.   Update global variational parameters $\mu_{q(\beta)}$, $\Sigma_{\beta}^q$ and $S_{q(D)}$ using  \\ [1mm]
\>\>\>  $\Sigma_{q(\beta)} \leftarrow \left[ (1-a_t)\Sigma_{q(\beta)}^{-1} +a_t \big\{ \Sigma_\beta^{-1} + \frac{n}{|B|} \sum_{i \in B}\big( \nu_{q(D)}  {\tilde{W}_i}^TS_{q(D)}^{-1}\tilde{W}_i  + V_i^T F_i V_i \big) \big\} \right]^{-1}$  \\ [1mm]
\>\>\>  $\mu_{q(\beta)} \leftarrow$ \= $\mu_{q(\beta)} + a_t \Sigma_{q(\beta)}\; \Big[ \frac{n}{|B|} \sum_{i \in B}\big\{ \nu_{q(D)}{\tilde{W}_i}^TS_{q(D)}^{-1}(\mu_{q(\tilde{\alpha}_i)}-\tilde{W}_i\mu_{q(\beta)}) + V_i^T(y_i-g_i) \big\}$\\ [1mm]
\>\>\>\> $ -\Sigma_\beta^{-1}\mu_{q(\beta)}\Big]$  \\ [1mm]
\>\>\> $S_{q(D)} \leftarrow$ \= $(1-a_t)S_{q(D)}+a_t\Big[\frac{n}{|B|} \sum_{i \in B} \big\{ (\mu_{q(\tilde{\alpha}_i)}-\tilde{W}_i \mu_{q(\beta)}) (\mu_{q(\tilde{\alpha}_i)}-\tilde{W}_i \mu_{q(\beta)})^T$   \\ [1mm] 
\>\>\>\>  $+ \Sigma_{q(\tilde{\alpha}_i)}+\tilde{W}_i\Sigma_{q(\beta)}\tilde{W}_i^T \big\}+S\Big] $ \\ [-4mm]
\\[-9mm]
\end{tabbing}\hrule}
\caption{\label{Alg 2} Stochastic nonconjugate variational message passing for GLMMs.}
\end{Algorithm}

To initialize Algorithm 2, we recommend using the fit from penalized quasi-likelihood, which can be implemented in {\ttfamily R} via the function {\ttfamily glmmPQL} in the package {\ttfamily MASS} \citep{Venables2002}. Alternatively, for large data sets where penalized quasi-likelihood converges too slowly, we can use the fit from the GLM (obtained by pooling all data and setting random effects as zero) for initialization. For instance, we can set $\mu_{q(\beta)}$ and $\Sigma_{q(\beta)}$ respectively as estimates of the regression coefficients and their covariances from the GLM, $S_{q(D)}=(\nu_{q(D)}-r-1)\hat{R}$ where $\nu_{q(D)}=\nu+n$, $\mu_{q(\tilde{\alpha}_i)}=\tilde{W}_i\mu_{q(\beta)}$ and $\Sigma_{q(\tilde{\alpha}_i)} =\hat{R}$. \citet{Kass2006} gave a justification of $\hat{R}$ [defined in \eqref{Kassprior}] being a reasonable guess for $D$ in the absence of any other prior knowledge. The tuning parameters $W_i$ can be initialized by setting $D=\hat{R}$ and $\eta_i=X_i \mu_{q(\beta)}$ (for logistic GLMMs). 

In step 1, mini-batches may be selected with or without replacement from the whole data set. Here, we consider sampling randomly at each iteration without replacement. Suppose the data set consist of $n$ clusters and we randomly select $|B|$ clusters at the first iteration. At the second iteration, we will randomly sample $|B|$ clusters from the remaining $n-|B|$ clusters and so on. Algorithm 2 is considered to have made a sweep through the data when all clusters have been used once. This process is then repeated. Mini-batches in each sweep are sampled randomly and do not depend on those in previous sweeps. We allow mini-batches in each sweep to differ in size by one when $n$ is not divisible by $|B|$. The advantage of sampling without replacement is that this scheme ensures all clusters (and local variational parameters) have been used or updated once in each sweep.

In step 2, we consider a loose criterion for assessing convergence to reduce computational overhead. Suppose mini-batch $B$ consist of units $\{j_1,\dots,j_{|B|}\}$. We define $\mu_{q(\tilde{\alpha})}^B = [\mu_{q(\tilde{\alpha}_{j_1})}^T, \dots, \mu_{q(\tilde{\alpha}_{j_{|B|}})}^T]^T$ and terminate repetitions in step 2 when $\frac{ \lVert {\mu_{q(\tilde{\alpha})}^B}\negthinspace^{(t)} - {\mu_{q(\tilde{\alpha})}^B}\negthinspace^{(t-1)} \lVert }{ \lVert {\mu_{q(\tilde{\alpha})}^B}\negthinspace^{(t)} \lVert} < 0.05$, where $\parallel \cdot \parallel$ represents the Euclidean norm. Typically 3--7 repetitions are required for each mini-batch in the first sweep. The number of repetitions reduces steadily with the number of sweeps and usually just a single update is required by the third sweep.

For the examples in this paper, we did not update the tuning parameters $W_i$ beyond initialization when the partially noncentered parametrization was used. While updating tuning parameters (at the end of each cycle in Algorithm \ref{Alg 1} or at the end of each sweep in Algorithm \ref{Alg 2}) can lead to further improvements, more computation is required and this can be time-consuming for large data sets. A good initialization of the tuning parameters based on say penalized quasi-likelihood usually suffices. 

The choice of step sizes $a_t$ can strongly influence the performance of a stochastic approximation algorithm \citep{Jank2006}. We discuss the choice of a gain sequence for Algorithm \ref{Alg 2} in the next section.

\subsection{Switching from stochastic to standard version}
Determining an appropriate stopping criterion for a stochastic approximation algorithm can be challenging. Some commonly used criteria include stopping when the relative change in parameter values or objective function is sufficiently small or when the gradient of the objective function is sufficiently close to zero \citep{Spall2003}. Such criteria do not provide any guarantees of the terminal iterate being close to the optimum however, and may be satisfied by random chance. \citet{Booth1999} recommend applying such rules for several consecutive iterations to minimize chances of a premature stop. However, \citet{Jank2006} gave an illustrative example to show that even this may not be enough of a safeguard. Moreover, stochastic approximation can become excruciatingly slow in later iterations due to small step sizes.

\begin{figure}[t]
\centering
\includegraphics[width=0.9\textwidth]{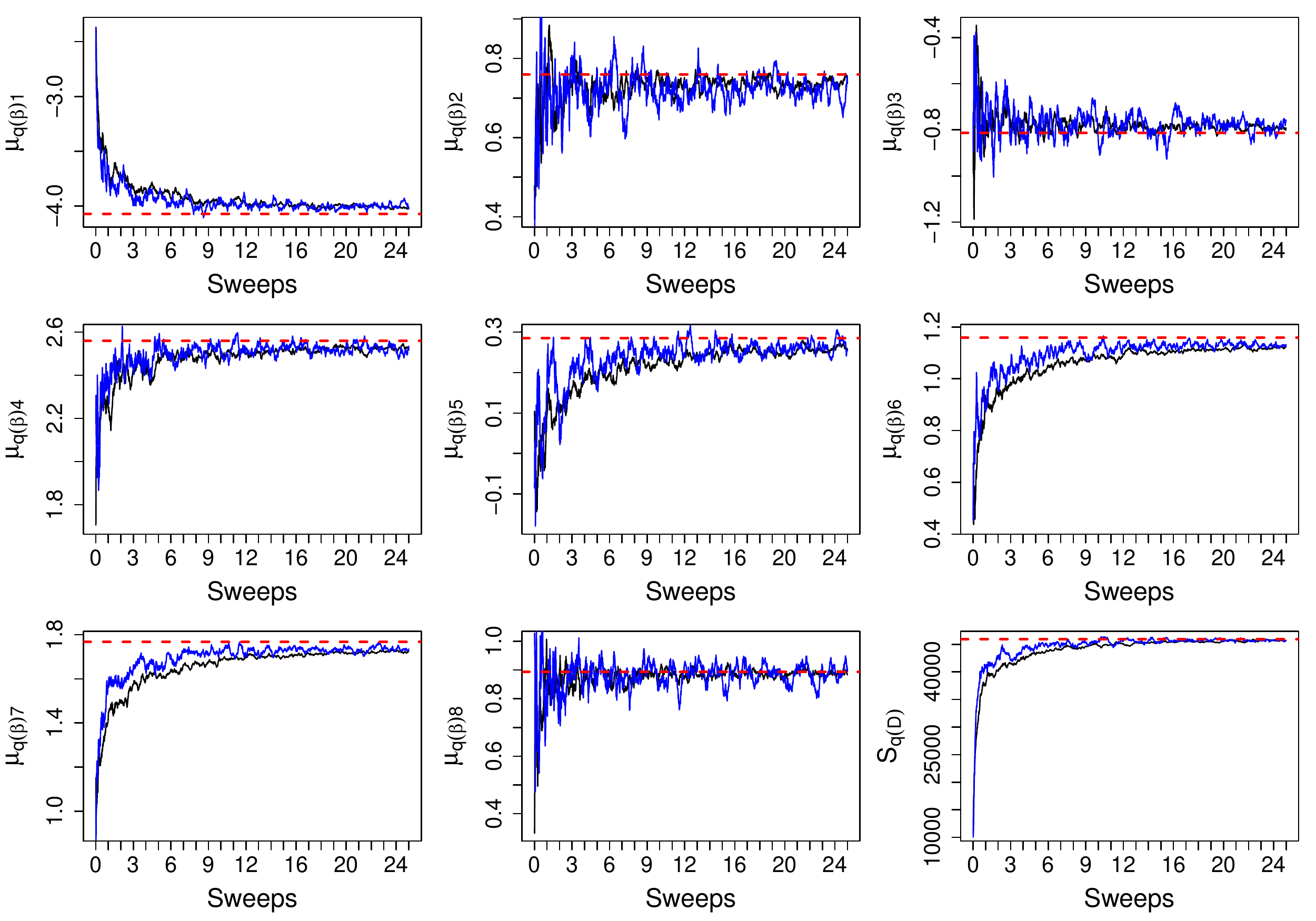}
\caption{\label{global} Polypharmacy simulated data ($n=10000$). Global variational parameters $\mu_{q(\beta)}$ and $S_{q(D)}$ fitted using Algorithm 2 plotted against number of sweeps. Mini-batches $|B|=100$ and step size $a_t = 1/(t+1)^\alpha$. Blue trajectories correspond to $\alpha=0.55$ and black to $\alpha=0.65$. Red dotted line denotes values obtained using Algorithm 1.}
\end{figure}

Our experimentations with moderate-sized data sets indicate that gains made by Algorithm \ref{Alg 2} are usually largest in the first few sweeps. However, beyond a certain point, it can become slower than Algorithm \ref{Alg 1} if step sizes are too small or iterates simply bounce around if step sizes are still too big. An example is shown in Figure \ref{global} where global variational parameters $\mu_{q(\beta)}$ and $S_{q(D)}$ are plotted against iterations $t$ (or number of sweeps). Here Algorithm 2 is applied to a simulated data set of size $n=10000$ (details in Example \ref{poly}) and mini-batches of size $|B|=100$ are used with step size $a_t = 1/(t+1)^\alpha$. Blue trajectories correspond to $\alpha=0.55$ and black to $\alpha=0.65$. Red dotted lines represent values obtained using Algorithm 1. Figure \ref{global} shows that the blue and black trajectories converge towards the red dotted lines quickly in the first few sweeps. However, full convergence takes much longer. A larger step size ($\alpha=0.55$) implies faster convergence at first but the iterates bounce around the optimum eventually if step sizes are still too large. A possible remedy to this is iterate averaging \citep{Polyak1992}.

We suggest switching to Algorithm \ref{Alg 1} when Algorithm \ref{Alg 2} shows signs of slowing down. Using the lower bound both as a switching and stopping criterion, we propose switching from stochastic to standard nonconjugate variational message passing when the relative increase in the lower bound after a sweep is less than $10^{-3}$, and terminating Algorithm \ref{Alg 1} when the relative increase in the lower bound is less than $10^{-6}$. For large datasets or streaming data, it might be more practical to terminate Algorithm \ref{Alg 2} beyond a certain period of available runtime. To switch from Algorithm \ref{Alg 2} to \ref{Alg 1}, the final setting of local and global variational parameters computed using Algorithm \ref{Alg 2} is used as initialization of Algorithm \ref{Alg 1}.

Let $M$ denote the number of iterations required to make a sweep through the data set. Following \cite{Spall2003}, we consider step sizes of the form $a_t=\frac{1}{t+A}$ setting $a=1$ and $\alpha=1$ in \eqref{Spallstepsize}. We let $t = s_w + \frac{m}{M}$ where $0 \leq m \leq M-1$ denotes the number of mini-batches that has been analysed at the $s_w$th sweep. This specification slows down the rate of decrease in step size within each sweep and the larger step sizes help iterates move faster towards the optimum. We investigate performance of different stability constants $A$ for various mini-batch sizes. Smaller values of $\alpha$ correspond to a slower decrease in step size and are desirable in some cases as they provide bigger step sizes in later iterations. For our proposed strategy, we observed that smaller mini-batch sizes generally performed better. Since smaller step sizes are preferred for smaller mini-batch sizes \citep[see][]{Hoffman2010}, we set $\alpha=1$ for simplicity and report results only for this case.

Recently, \cite{Ranganath2013} developed an adaptive learning rate for stochastic variational inference, which is designed to minimize the expected distance between stochastic and optimal updates of the global variational parameters. They showed that adaptive step sizes led to improved convergence for the latent Dirichlet allocation model in topic modelling. It might be possible to extend this adaptive learning rate to nonconjugate models and we are working on this area. A wide variety of approaches have been developed to enhance the rate of convergence of stochastic approximation algorithms, and examples include iterate averaging \citep{Polyak1992}, momentum method \citep{Tseng1998} and gradient averaging \citep{Xiao2010}. See \cite{Roux2012} for the stochastic average gradient method as well as a review of other approaches.

\section{Prior-likelihood conflict diagnostics as a by-product of variational message passing}
In this section, we consider diagnostic tests for identifying divergent units in GLMMs. Such diagnostics are useful for detecting institutions (e.g. hospitals, trusts or schools) which deviate from the rest in a certain outcome. In healthcare for instance, it may be of interest to identify hospitals which are divergent in terms of quality of care provided or choice of surgical procedure for treating a cancer \citep{Farrell2010}. We demonstrate how prior-likelihood conflict diagnostics for identifying divergent units can be obtained as a by-product of nonconjugate variational message passing. The intuitive idea is that messages coming from above and below a node in a hierarchical model can be separated and ``mixed messages"  indicate conflict. Our ``mixed messages" diagnostics can be shown to approximate the conflict diagnostics of \citet{Marshall2007}. We start with a review of the simulation-based diagnostic test \citep{Marshall2007}, which is based on measuring the conflict between likelihood of a parameter and its predictive prior given the remaining data. Subsequently, we show how their approach can be approximated in the variational message passing framework. 

\subsection{Cross-validatory conflict $p$-values from a simulation-based approach} \label{CV}
For GLMMs with a partially noncentered parametrization, the linear predictor is 
\begin{equation*}
\eta_i=V_i \beta + Z_i \tilde{\alpha}_i \;\; \text{where} \;\; \tilde{\alpha}_i \sim N(\tilde{W}_i \beta, D) \;\; \text{for} \;\; i=1,\dots,n.
\end{equation*}
To identify units that do not appear to be drawn from the assumed random effects distributions, \citet{Marshall2007} suggest comparing replicates of $\tilde{\alpha}_i$ from its likelihood and predictive prior. A predictive prior replicate $\tilde{\alpha}_i^ {\text{rep}}$ is first generated from 
\begin{equation}\label{rep}
p_r(\tilde{\alpha}_i|y_{-i})=\int p(\tilde{\alpha}_i|\beta,D)\,p(\beta,D|y_{-i}) \;d\beta\, d D
\end{equation}
where $y_{-i}$ denotes observed data $y$ with unit $i$ left out. This replicate can be obtained by generating $\beta^ {\text{rep}}$ and $D^ {\text{rep}}$ from $p(\beta,D|y_{-i})$ using MCMC, followed by simulation of $\tilde{\alpha}_i^ {\text{rep}}|\beta^ {\text{rep}},D^ {\text{rep}}$. A likelihood replicate $\tilde{\alpha}_i^ {\text{lik}} \sim p(\tilde{\alpha}_i|y_i)$ is then generated using only the unit $y_i$ being tested and a non-informative prior $p(\tilde{\alpha}_i)$ for $\tilde{\alpha}_i$. \cite{Marshall2007} recommend using the Jeffreys's prior as a noninformative prior for $\tilde{\alpha}_i$ \citep[see][]{Box1973}. These prior and likelihood replications represent independent sources of evidence about $\tilde{\alpha}_i$ and conflict between them suggests discrepancies in the model.

The discussion above ignores nuisance parameters. For GLMMs, we need to regard $\beta$ as a nuisance parameter. As $p(\tilde{\alpha}_i|y_i) \propto p(\tilde{\alpha}_i) \int p(y_i|\beta,\tilde{\alpha}_i)\,p(\beta|\tilde{\alpha}_i) \, d\beta$ and $\beta$ is not estimable from individual unit $i$, \citet{Marshall2007}[pg. 420] recommend generating $\tilde{\alpha}_i^ {\text{lik}}$ from
\begin{equation*}
p_l(\alpha_i|y) \propto p(\tilde{\alpha}_i)\int p(y_i|\tilde{\alpha}_i,\beta)p(\beta|y_{-i})\,d\beta.
\end{equation*}
Note that the two replications $\tilde{\alpha}_i^ {\text{rep}}$ and $\tilde{\alpha}_i^ {\text{lik}}$ are no longer entirely independent as $y_{-i}$ will slightly influence $\tilde{\alpha}_i^ {\text{lik}}$ through $\beta$.

To compare prior and likelihood replicates, \citet{Marshall2007} considered $\tilde{\alpha}_i^ {\text{diff}}=\tilde{\alpha}_i^ {\text{rep}}-\tilde{\alpha}_i^ {\text{lik}}$ and calculated a conflict $p$-value,
\vspace{-2mm}
\begin{equation*}
p_{i,\text{con}}^l =P(\tilde{\alpha}_i^ {\text{diff}} \leq 0|y)
\end{equation*}
as the proportion of times simulated values of $\tilde{\alpha}_i^ {\text{diff}}$ are less than or equal to zero for scalar $\tilde{\alpha}_i$. Depending on the context, the upper tail area $p_{i,\text{con}}^u=1-p_{i,\text{con}}^l$ or two-sided $p$-value $2 \times \text{min}(p_{i,\text{con}}^l,p_{i,\text{con}}^u)$ may be of interest instead. If $\tilde{\alpha}_i^ {\text{diff}}$ is not a scalar, 
\begin{equation*}
\Delta = \text{E}(\tilde{\alpha}_i^ {\text{diff}}|y)^T \text{Cov}(\tilde{\alpha}_i^ {\text{diff}}|y)^{-1} \text{E}(\tilde{\alpha}_i^ {\text{diff}}|y)
\end{equation*} 
can be used as a standardized discrepancy measure. If we further assume a multivariate normal distribution for $\tilde{\alpha}_i^ {\text{diff}}$, then a conflict $p$-value for testing $\tilde{\alpha}_i^ {\text{diff}} = 0$ can be calculated as $P(\chi_r^2 > \Delta)$, where $\chi_r^2$ denotes a Chi-square random variable with $r$ degrees of freedom. Further discussion on $p$-values in multivariate case can be found in \cite{Presanis2013}.

As MCMC methods are not well-suited to cross-validation approaches, \cite{Marshall2007} proposed an alternative simulation-based full-data approach. The procedure is the same as before except that $\tilde{\alpha}_i^ {\text{rep}}|\beta^ {\text{rep}},D^ {\text{rep}}$ is simulated using $\beta^ {\text{rep}}$, $D^ {\text{rep}}$ generated from $p(\beta,D|y)$, without leaving out $y_i$. Mild conservatism is introduced as $y_i$ will influence $\tilde{\alpha}_i^ {\text{rep}}$ slightly through $\beta$ and $D$.

\subsection{Conflict $p$-values from nonconjugate variational message passing}
Next, we show how approximate conflict $p$-values can be calculated within nonconjugate variational message passing. From \eqref{NCVMP_update}, the update for $\lambda_{\tilde{\alpha}_i}$ is
\begin{equation*}
\mathcal{V}_{\tilde{\alpha}_i}(\lambda_{\tilde{\alpha}_i})^{-1}\nabla_{\lambda_{\tilde{\alpha}_i}} E_q\{ \log p(\tilde{\alpha}_i| \beta, D) \} + \mathcal{V}_{\tilde{\alpha}_i}(\lambda_{\tilde{\alpha}_i})^{-1}\nabla_{\lambda_{\tilde{\alpha}_i}} E_q\{ \log p(y_i| \tilde{\alpha}_i, \beta) \}.
\end{equation*}
The first term can be considered as a message from the prior $p(\tilde{\alpha}_i|\beta,D)$ and the second term a message from the likelihood $p(y_i|\tilde{\alpha}_i,\beta)$ of unit $y_i$. We argue below that the first message from the prior can be interpreted as natural parameter of a Gaussian approximation say $N(\mu_{\text{rep}},\Sigma_{\text{rep}})$ to  $p_r(\tilde{\alpha}_i|y_{-i})$. On the other hand, the second message from the likelihood can be interpreted as natural parameter of a Gaussian approximation say $N(\mu_{\text{lik}},\Sigma_{\text{lik}})$ to $p_l(\tilde{\alpha}_i|y)$. This implies that $\tilde{\alpha}_i^ {\text{rep}} \sim N(\mu_{\text{rep}},\Sigma_{\text{rep}})$ and $\tilde{\alpha}_i^ {\text{lik}} \sim N(\mu_{\text{lik}},\Sigma_{\text{lik}})$. If we further assume $\tilde{\alpha}_i^ {\text{rep}}$ and $\tilde{\alpha}_i^ {\text{lik}}$ are independent, then $\tilde{\alpha}_i^ {\text{diff}} \sim N(\mu_{\text{rep}}-\mu_{\text{lik}}, \Sigma_{\text{rep}}+\Sigma_{\text{lik}})$. Even though $\tilde{\alpha}_i^ {\text{rep}}$ and $\tilde{\alpha}_i^ {\text{lik}}$ are not entirely independent, the dependence between $\tilde{\alpha}_i^ {\text{rep}}$ and $\tilde{\alpha}_i^ {\text{lik}}$ will be increasingly weak as the number of clusters increases. Since these messages are computed in the nonconjugate variational message passing algorithm, conflict $p$-values can be calculated easily at convergence for identifying divergent units. 

The arguments presented below are by no means rigorous. However, they lend some insight into how conflict $p$-values can be approximated from nonconjugate variational message passing and experimental results suggest the approximations work well in practice. For large data sets, automatic computation of diagnostics for prior-likelihood conflict can be an attractive alternative to simulation-based MCMC approaches. They are also useful generally as initial screening tools and clusters flagged as divergent can be studied more closely and possibly conflict $p$-values recomputed by Monte Carlo. 

First, consider the message from the prior. If we treat the message as natural parameter of a normal distribution, we get $\mu_{\text{rep}} = \tilde{W}_i \mu_{q(\beta)}$ and $\Sigma_{\text{rep}} = S_{q(D)}/\nu_{q(D)}$. For large data sets, $p(\beta,D|y_{-i})$ is close to $p(\beta,D|y)$ and we approximate $p(\beta,D|y_{-i})$ in \eqref{rep} by the variational posterior $q(\beta|\lambda_\beta)q(D|\lambda_D)$. This combined with Jensen's inequality gives
\begin{equation*}
\begin{aligned}
p_r(\tilde{\alpha}_i|y_{-i}) &\approx \int p(\tilde{\alpha}_i|\beta,D)\,q(\beta|\lambda_\beta)q(D|\lambda_D) \;d\beta\, d D \\
& \geq \exp[E_{-\tilde{\alpha}_i}\{\log p(\tilde{\alpha}_i|\beta,D)\}].
\end{aligned}
\end{equation*}
While $\exp[E_{-\tilde{\alpha}_i}\{\log p(\tilde{\alpha}_i|\beta,D)\}]$ is only a lower bound to $p_r(\tilde{\alpha}_i|y_{-i})$, we find that by using it as an approximation to $p_r(\tilde{\alpha}_i|y_{-i})$, we get $p_r(\tilde{\alpha}_i|y_{-i}) \propto \exp[E_{-\tilde{\alpha}_i}\{\log p(\tilde{\alpha}_i|\beta,D)\}]$. This gives $\tilde{\alpha}_i^ {\text{rep}} \sim N(\tilde{W}_i \mu_{q(\beta)}, S_{q(D)}/\nu_{q(D)})$, which is what we would get if we interpret the first message as being the natural parameter of a Gaussian approximation to $p_r(\tilde{\alpha}_i|y_{-i})$.

Next, consider the second message from the likelihood. If we treat the message as the natural parameter of a normal distribution, it can be shown that $\Sigma_{\text{lik}}^{-1} = Z_i^T F_i Z_i$ and $\mu_{\text{lik}} =\mu_{q(\tilde{\alpha}_i)}+\Sigma_{\text{lik}} Z_i^T (y_i-g_i)$. Now consider the sum of the two messages. This gives us the natural parameter of $q(\tilde{\alpha}_i|\lambda_{\tilde{\alpha}_i})$ which is an approximation of $p(\tilde{\alpha}_i|y)$. Note that 
\begin{equation*}
\Sigma_{\text{rep}}^{-1} + \Sigma_{\text{lik}}^{-1} = \Sigma_ {q(\tilde{\alpha}_i)}^{-1} \;\;\text{and}\;\; \Sigma_{\text{rep}}^{-1} \mu_{\text{rep}} + \Sigma_{\text{lik}}^{-1} \mu_{\text{lik}} = \Sigma_ {q(\tilde{\alpha}_i)}^{-1} \mu_ {q(\tilde{\alpha}_i)}.
\end{equation*}
If we think of $p(\tilde{\alpha}_i|y_{-i})$ as the `prior' to be updated when $y_i$ becomes available, we have 
\begin{equation*}
p(\tilde{\alpha}_i|y) \propto p(\tilde{\alpha}_i|y_{-i})p(y_i|\tilde{\alpha}_i,y_{-i}) \Rightarrow \frac{p(\tilde{\alpha}_i|y)}{p(\tilde{\alpha}_i|y_{-i})} \propto p(y_i|\tilde{\alpha}_i,y_{-i}). 
\end{equation*}
Interpreting the first message as a Gaussian approximation to $p(\tilde{\alpha}_i|y_{-i})$ and the sum of the two messages as a Gaussian approximation to $p(\tilde{\alpha}_i|y)$, the ratio of these two normal distributions gives an approximation (up to a proportionality constant) of $p(y_i|\tilde{\alpha}_i,y_{-i})$. As a function of $\tilde{\alpha}_i$, the ratio of the two normal distributions is proportional to 
\begin{equation*}
\frac{\exp\{-\frac{1}{2}(\tilde{\alpha}_i-\mu_{q(\tilde{\alpha}_i)})^T \Sigma_{q(\tilde{\alpha}_i)}^{-1}(\tilde{\alpha}_i-\mu_{q(\tilde{\alpha}_i)})\}}{\exp\{-\frac{1}{2}(\tilde{\alpha}_i-\mu_{\text{rep}})^T \Sigma_{\text{rep}}^{-1}(\tilde{\alpha}_i-\mu_{\text{rep}})\}},
\end{equation*}
which gives a normal distribution with mean $\mu_{\text{lik}}$ and covariance $\Sigma_{\text{lik}}$, precisely that given by the second message. As
\begin{equation*}
p(y_i|\tilde{\alpha}_i,y_{-i})=\int p(y_i|\beta,\tilde{\alpha}_i)p(\beta|\tilde{\alpha}_i,y_{-i})\,d\beta
\end{equation*}
and $p(\beta|\tilde{\alpha}_i,y_{-i})$ is close to $p(\beta|y_{-i})$ when the number of clusters is large (in the sense that dependence of $\beta$ on $\tilde{\alpha}_i$ is reduced), the second message can be considered as the natural parameter of a Gaussian approximation to $p_l(\tilde{\alpha}_i|y)$ if we assume a uniform prior for $p(\tilde{\alpha}_i)$. The arguments above generalize to detecting conflict for other parameters of the model as well.

While the discussion here uses the partially noncentered parametrization, conclusions hold for the centered and noncentered parametrizations as well. We observed small differences in conflict $p$-values computed using different parametrizations,  which is due likely to varying accuracy of approximations to the true posterior. To compare the accuracy of different approaches, we first transform the conflict $p$-values to $z$-scores to reflect the importance of good agreement at the extremes \citep{Marshall2007}. Using the cross-validatory conflict $p$-values as a ``gold-standard", we use the mean absolute difference in $z$-scores,
\begin{equation*}
\frac{1}{n} \sum_{i=1}^n |\Phi^{-1}(p_{i,\text{con}}^\text{CV}) - \Phi^{-1}(p_{i,\text{con}}^\text{method})|,
\end{equation*}
as a measure of the degree of agreement between the cross-validatory conflict $p$-values ($p_{i,\text{con}}^\text{CV}$) and conflict $p$-values computed from the method we are trying to assess ($p_{i,\text{con}}^\text{method}$). 

To compute conflict-$p$ values for large data sets, one needs to ensure that local variational parameters for every unit are optimized. As Algorithm 2 focuses on optimization of global variational parameters using stochastic approximation, not all local variational parameters may have been fully optimized when the global variational parameters have converged. This can be resolved by performing an additional step of optimizing local variational parameters for every unit as a function of the converged global variational parameters. Alternatively, our proposed strategy of switching from Algorithm 2 to 1 also ensures that local variational parameters for every unit are optimized. However, due to the difficulty in computing conflict $p$-values for large data sets using cross-validatory or even full-data approaches with MCMC, we focus on comparisons with nonconjugate variational message passing using only small data problems in the examples.

\section{Examples} \label{eg}
In sections \ref{Bristol} and \ref{epilepsy}, we use the Bristol inquiry data and epilepsy data to compare conflict $p$-values computed using nonconjugate variational message passing with those obtained using the simulation-based cross-validatory approach of \cite{Marshall2007}. An additional example on Madras schizophrenia data can be found in Appendix \ref{schizo}. These data sets are relatively small and we only use Algorithm \ref{Alg 1} for fitting. 

In sections \ref{poly} and \ref{skincancer}, we use moderately large simulated data sets to illustrate the improvements in efficiency that can be obtained by using stochastic nonconjugate variational message passing in the initial stage of optimization. We compare performances of Algorithms \ref{Alg 1} and \ref{Alg 2} for the simulated data sets using only the partially noncentered parametrization. Algorithms \ref{Alg 1} and \ref{Alg 2} were initialized using penalized quasi-likelihood in all examples except for the large simulated data set in Section \ref{skincancer}, where penalized quasi-likelihood converges too slowly. The GLM fit was used instead for initialization.

In all examples, fitting via MCMC was performed in OpenBUGS \citep{Lunn2009} through {\ttfamily R} by using {\ttfamily R2OpenBUGS} as an interface. {\ttfamily R2OpenBUGS} was adapted by Neal Thomas from {\ttfamily R2WinBUGS} \citep{Sturtz2005}. The MCMC algorithm was initialized using penalized quasi-likelihood and the same priors were used in MCMC and nonconjugate variational message passing. We consider a vague $N(0,1000)$ prior for $\beta$ in each case. All code was written in {\ttfamily R} and run on a dual processor Windows PC 3.30 GHz workstation. Computation times reported are in seconds (s).

In some examples below, the variational posterior approximations are biased as compared to results from MCMC. This is due to the assumption of a factorized variational posterior and the impact of this restriction depends on how strong posterior dependence is among the factored variables. In VB, the posterior variance tends to be underestimated and this issue has been noted by \cite{Wang2005} and \cite{Bishop2006}. Recently, \cite{Zhao2013} proposed some diagnostics for assessing how well VB approximates the true posterior as well as correction measures that can be undertaken when the approximation error is large. \cite{Salimans2013} developed stochastic approximation methods for hierarchical approximations that allow independence assumptions in VB to be relaxed.

\subsection{Bristol inquiry data} \label{Bristol}
In 1998, a public inquiry was set up to look into the management of children receiving complex cardiac surgical services at the Bristol Royal Infirmary. The outcomes of surgical services at Bristol, UK, relative to other specialist centres was a key issue. We consider a subset of the data recorded by Hospital Episode Statistics on mortality rates in open surgeries for 12 hospitals including Bristol (hospital 1), for children under 1 year old, from 1991 to 1995 \citep[see][Table 1]{Marshall2007}. Although the number of clusters is small in this example whereas our methodology is motivated by applications to large data sets, this example is interesting as a benchmark data set in the literature for computing conflict diagnostics using nonconjugate variational message passing. 

Let $y_{ij} \sim \text{Bernoulli}(\pi_i)$ where $y_{ij}=1$ if patient $j$ at hospital $i$ died and 0 otherwise. We use $Y_i=\sum_{j=1}^{n_i} y_{ij}$ to denote the number of deaths at hospital $i$, $i=1,\dots,12$. Let
\begin{equation*}
\text{logit}(\pi_i)=\beta+u_i\;\;\;\text{where}\;\;\;u_i \sim N(0,D).
\end{equation*}
In the cross-validatory approach, each hospital $i$ was removed in turn from the analysis, and $\beta^ {\text{rep}}, D^ {\text{rep}}|y_{-i}$ were generated using MCMC followed by a simulated $\pi_i^ {\text{rep}}|\beta^ {\text{rep}},D^ {\text{rep}}$. Assuming a Jeffreys's prior for $\pi_i$, a $\pi_i^{\text{lik}}$ was simulated from $ p(\pi_i|y_i) = \text{Beta}(Y_i+0.5,n_i-Y_i+0.5)$. Excess mortality is of concern and the upper-tail area is used as a 1-sided $p$-value so that $p_{i,\text{con}}=P(\pi_i^ {\text{rep}} \geq \pi_i^{\text{lik}})$. For each fitting via MCMC, two chains were run simultaneously to assess convergence, each with 51,000 iterations, and the first 1000 iterations were discarded in each chain as burn-in. Cross-validatory conflict $p$-values were calculated based on the remaining 100,000 simulations. The total time taken for model updating in OpenBUGS is 5 s $\times$ 12 = 60 s for the cross-validatory approach.

\begin{figure}[H]
\centering
\includegraphics[width=0.55\textwidth]{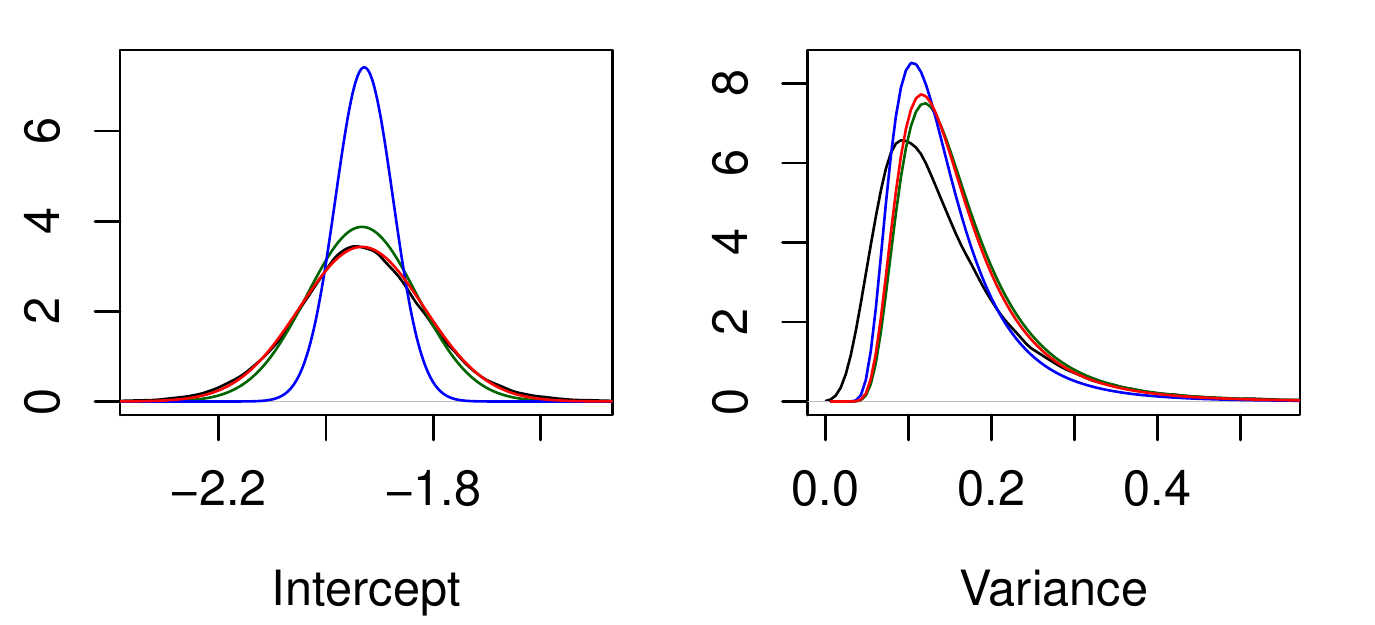}
\caption{\label{Bristolplot} Bristol data. Marginal posteriors estimated by MCMC (black) and Algorithm \ref{Alg 1} using the centered (green), noncentered (blue) and partially noncentered (red) parametrizations.}
\end{figure}

The variational lower bounds and CPU times taken for model fitting and computation of conflict $p$-values by Algorithm \ref{Alg 1} (via different parametrizations) and MCMC (full-data approach) are shown in Table \ref{Bristoltable}. Figure \ref{Bristolplot} shows the marginal posteriors of $\beta$ and $D$ estimated using MCMC and Algorithm \ref{Alg 1}. The partially noncentered parametrization attained the highest lower bound, was quick to converge and produced posterior approximations very close to that of MCMC.

\begin{table}[H]
\centering { \footnotesize \begin{tabular}{@{}lccKK@{}}  \hline
& noncentered  & centered  & partially noncentered & MCMC (full-data)  \\  \hline
Lower bound ($\mathcal{L}$) & -1213.7 & -1213.0 & -1212.9 & -- \\
Time (model fitting) & 7.6  & 3.7 & 3.8 & 5     \\ 
Time (computing conflict $p$-values) & 0.3 & 0.3 & 0.3 & 14.4 \\
Mean absolute difference in $z$-scores & 0.087 & 0.086 & 0.083 & 0.125 \\
\hline
\end{tabular}}
\caption{\label{Bristoltable} Bristol data. Variational lower bounds (first row), CPU times (s) for model fitting (second row) and computing conflict $p$-values (third row) and mean absolute difference in $z$-scores relative to cross-validatory approach (third row) for Algorithm \ref{Alg 1} (different parametrizations) and MCMC (full-data).}
\end{table}
\vspace{-6mm}
\vspace{-4mm}
\begin{figure}[H]
\centering
\begin{minipage}[c]{0.38\textwidth}
\centering
{\footnotesize
\begin{tabular}{*{3}{c}}
\hline hospital & $p_{i,\text{con}}^\text{CV}$ & $p_{i,\text{con}}^\text{NCVMP}$ \\ \hline
1 & 0.001 & 0.005\\
2 & 0.436 & 0.450\\
3 & 0.935 & 0.928\\
4 & 0.125 & 0.138\\
5 & 0.298 & 0.311\\
6 & 0.720 & 0.725\\
7 & 0.737 & 0.745\\
8 & 0.661 & 0.667\\
9 & 0.440 & 0.453\\
10 & 0.380 & 0.390\\
11 & 0.763 & 0.764\\
12 & 0.721 & 0.727\\
\hline
\end{tabular}}
\end{minipage}
\begin{minipage}[c]{0.58\textwidth}
\centering
\includegraphics[width=75mm]{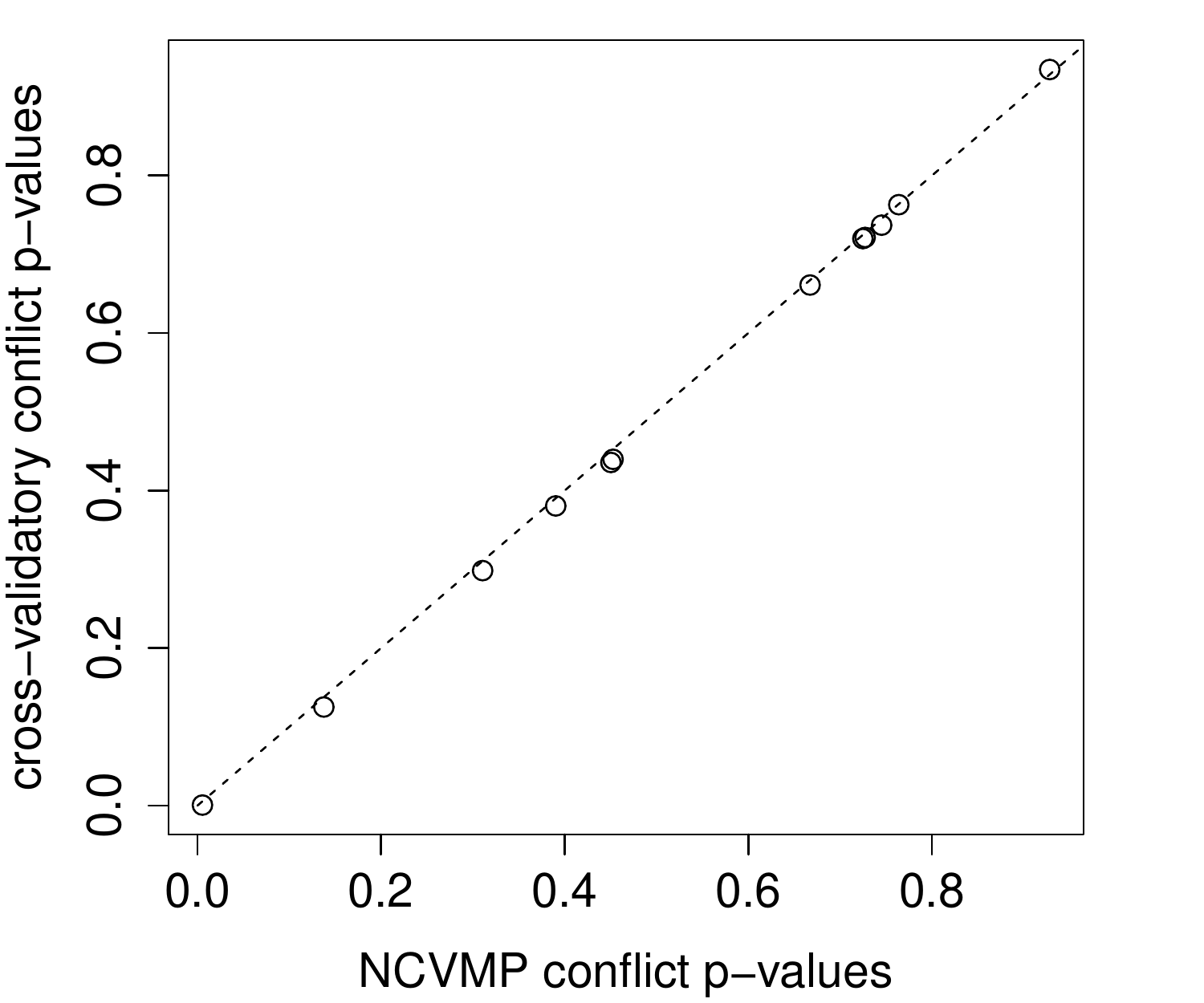}
\end{minipage}
\caption{ \label{bristolfig} Bristol data. Cross-validatory conflict $p$-values ($p_{i,\text{con}}^\text{CV}$) and conflict $p$-values from nonconjugate variational message passing ($p_{i,\text{con}}^\text{NCVMP}$) using a partially noncentered parametrization.}
\end{figure}

Figure \ref{bristolfig} compares conflict $p$-values computed using the cross-validatory approach and nonconjugate variational message passing using the partially noncentered parametrization. The plot indicates very good agreement between the two sets of $p$-values. Both approaches suggest hospital 1 (Bristol) is discrepant. The mean absolute difference in $z$-scores for nonconjugate variational message passing and the simulation-based full-data approach relative to the cross-validatory approach are given in Table \ref{Bristoltable}. Nonconjugate variational message passing does better than the simulation-based full-data approach both in terms of $z$-scores and computation time. The difference in conflict $p$-values computed using different parametrizations is small.

For this example, nonconjugate variational message passing is of an order of magnitude faster than the cross-validatory approach. We will see in the next two examples that the reduction in computation time is even greater for larger data sets. There are some difficulties in comparing nonconjugate variational message passing and MCMC in this way as the time taken for the variational algorithm to converge depends on the initialization, stopping rule and the rate of convergence is problem-dependent. The updating time for MCMC is also problem-dependent and depends on the length of burn-in and number of sampling iterations. It is clear, however, that for large data sets, the variational approach is attractive as an alternative to MCMC methods for obtaining prior-likelihood conflict diagnostics or as an initial screening tool.

\subsection{Epilepsy data} \label{epilepsy}

The epilepsy data set of \cite{Thall1990} contains records from a clinical trial of 59 patients with epilepsy. Each patient was randomly administered a new anti-epileptic drug, progabide, (Trt=1) or a placebo (Trt=0) and the number of seizures during the two weeks before each of four successive clinic visits (Visit, coded as $\text{Visit}_1=-0.3$, $\text{Visit}_2=-0.1$, $\text{Visit}_3=0.1$ and $\text{Visit}_4=0.3$) was recorded. The number of seizures during the 8-week period prior to randomization was also noted. We consider the logarithm of $\frac{1}{4}$ the number of baseline seizures (Base) and the logarithm of the age of patient (Age) as covariates. We center the covariate Age at its mean to improve mixing in MCMC methods.

\cite{Breslow1993} considered a Poisson random intercept and slope model:
\begin{equation}\label{epilmmodel}
\log \mu_{ij}= \beta_0+\beta_1\text{Base}_i+\beta_2\text{Trt}_i+ \beta_3 \text{Base}_i \times \text{Trt}_i+\beta_4 \text{Age}_i +\beta_5 \text{Visit}_{ij} +u_{1i}+u_{2i} \text{Visit}_{ij},
\end{equation}
for $i=1,\dots,59$, $j=1,\dots,4$ and $\left[\begin{smallmatrix} u_{1i}\\u_{2i} \end{smallmatrix}\right]\sim N\left(0,\left[\begin{smallmatrix} \sigma_{11}^2 & \sigma_{12}\\\sigma_{21} & \sigma_{22}^2 \end{smallmatrix}\right]\right)$. We compare conflict $p$-values computed using the cross-validatory approach and nonconjugate variational message passing for two models. Model I is a random intercept model where the random slope is dropped from \eqref{epilmmodel}. Model II is the random intercept and slope model in \eqref{epilmmodel}. We examine the suitability of the assumed random effects distribution and report two-sided conflict $p$-values for both models.

For simulation-based approaches, it is easier to work with the centered parametrization as handling of nuisance parameters is minimized (see details in Appendix C). Under this parametrization, there are no nuisance parameters in Model II and only $\beta_5$ needs to be regarded as a nuisance parameter in Model I. Each patient was removed in turn from the analysis in the cross-validatory approach. For each model fitting via MCMC, two chains were run simultaneously to assess convergence, each with 26,000 iterations, and the first 1000 iterations were discarded in each chain as burn-in. Cross-validatory conflict $p$-values were calculated based on the remaining 50,000 simulations. The total time taken for model updating in OpenBUGS is 61 s $\times$ 59 = 3599 s for Model I and 54 s $\times$ 59 = 3186 s for Model II. Simulation of prior and likelihood replicates of the centered random effects $\alpha_i$ was performed in {\ttfamily R}. To simulate likelihood replicates, we assume Jeffreys's prior for $\alpha_i$ and use adaptive rejection metropolis sampling via the {\ttfamily arms} function in the {\ttfamily HI} package \citep{Petris2003}.

\begin{table}[H]
\centering
{\footnotesize
\begin{tabular}{@{}lccKK@{}}
\hline
& noncentered  & centered  & partially noncentered & MCMC (full-data)  \\  \hline
Model I  \\
Lower bounds ($\mathcal{L}$) & -707.0 & -701.5 & -701.1 & -- \\
Time (model fitting) & 1.4  & 0.2 & 0.2 & 62    \\ 
Time (computing conflict $p$-values) & $<0.05$ & $<0.05$ & $<0.05$ & 4278.2 \\
Mean absolute difference in $z$-scores & 0.167 & 0.159 & 0.155 & 0.103 \\ [1mm] \hline
Model II \\
Lower bounds ($\mathcal{L}$) & -701.4 & -696.1 & -695.3 & -- \\
Time (model fitting) & 1.3  & 0.5 & 0.5 & 55     \\ 
Time (computing conflict $p$-values) & $<0.05$  & $<0.05$  & $<0.05$  & 3109.6 \\
Mean absolute difference in $z$-scores & 0.105 & 0.107 & 0.101 & 0.116 \\
\hline
\end{tabular}}
\caption{\label{epilepsytable} Epilepsy data. Variational lower bounds (first row), CPU times (s) for model fitting (second row) and computing conflict $p$-values (third row), and mean absolute difference in $z$-scores relative to cross-validatory approach (third row) for Algorithm \ref{Alg 1} (different parametrizations) and MCMC (full-data).}
\end{table}
\begin{figure}[H]
\centering
\includegraphics[width=0.9\textwidth]{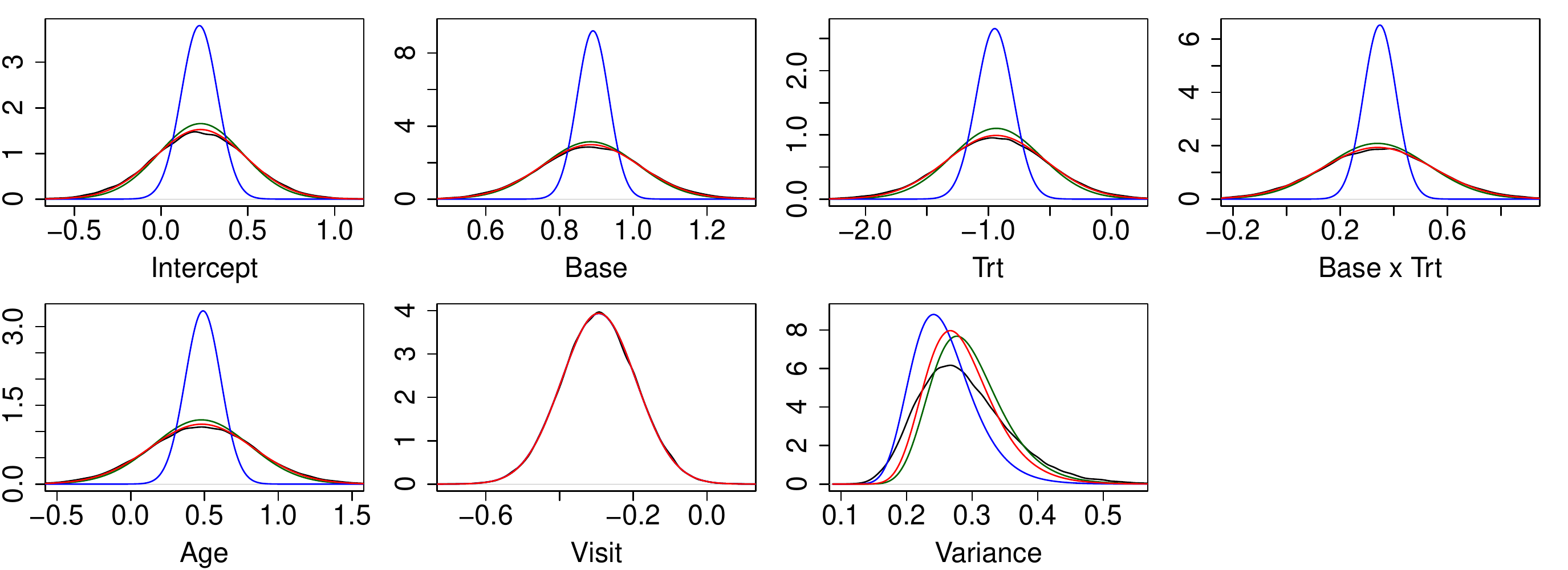}
\caption{\label{epilmar}Epilepsy data Model I. Marginal posteriors estimated by MCMC (black) and Algorithm \ref{Alg 1} using the centered (green), noncentered (blue) and partially noncentered (red) parametrizations.}
\end{figure}

Variational lower bounds and CPU times taken for model fitting and computation of conflict $p$-values by Algorithm \ref{Alg 1} (via different parametrizations) and MCMC (full-data approach) are given in Table \ref{epilepsytable}. Marginal posteriors of parameters in Model I estimated using MCMC and Algorithm \ref{Alg 1} are given in Figure \ref{epilmar}. Comparison of parameter estimates for Model II can be found in \cite{Tan2013}. The partially noncentered parametrization performed very well in posterior approximations and was quick to converge.

\begin{figure}[H]
\centering
\includegraphics[width=0.75\textwidth]{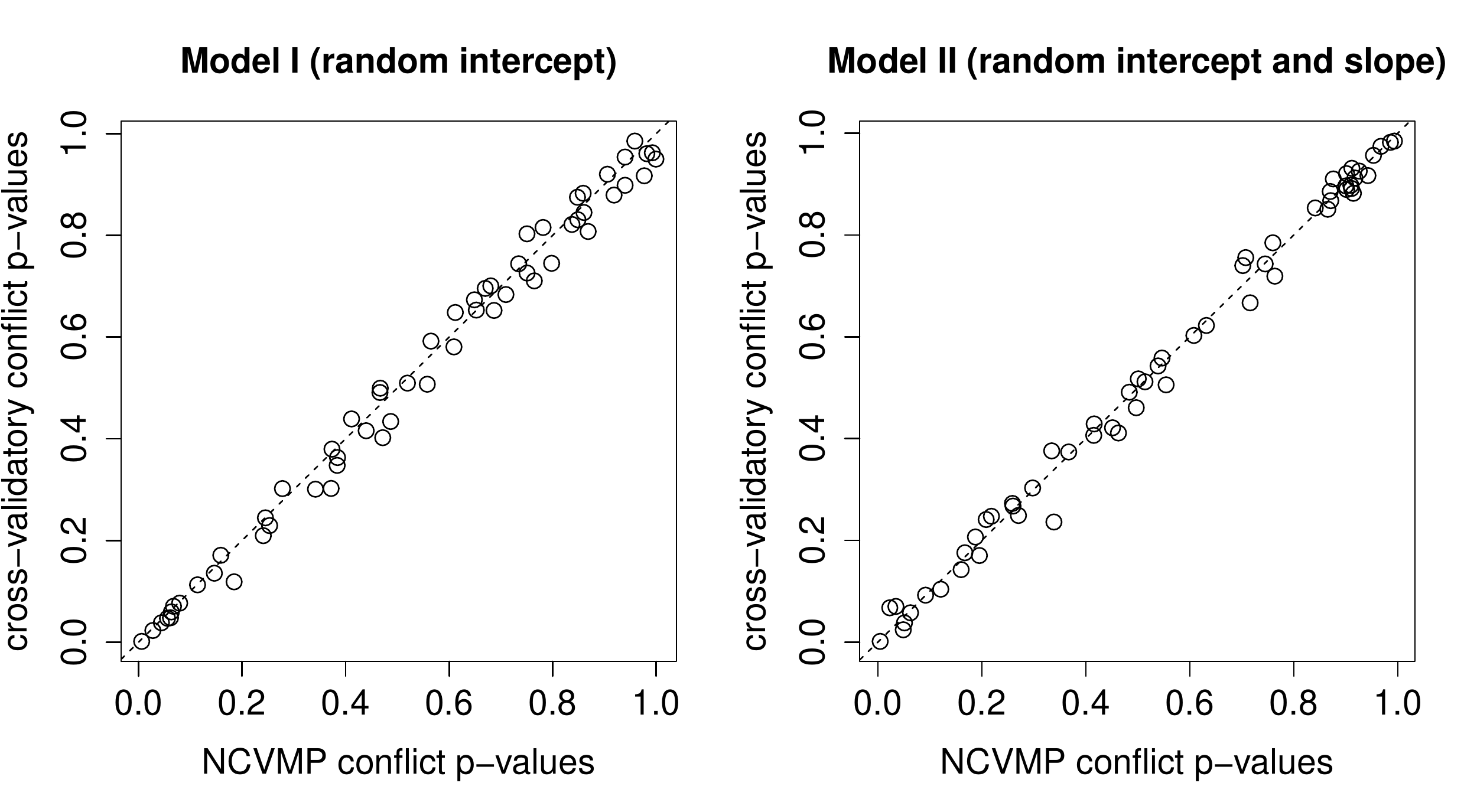} 
\vspace{-2mm}
\caption{\label{epilpvplot} Epilepsy data. Cross-validatory conflict $p$-values plotted against conflict $p$-values from nonconjugate variational message passing using a partially noncentered parametrization, for Model I (left) and Model II (right).}
\end{figure}

Cross-validatory conflict $p$-values are plotted against conflict $p$-values from nonconjugate variational message passing using the partially noncentered parametrization in Figure \ref{epilpvplot}, for Model I (left) and Model II (right). The mean absolute difference in $z$-scores for nonconjugate variational message passing and the simulation-based full-data approach relative to the cross-validatory approach are given in Table \ref{epilepsytable}. Figure \ref{epilpvplot} shows good agreement between cross-validatory conflict $p$-values and conflict $p$-values computed using nonconjugate variational message passing. The agreement is better in Model II and this is reflected in the $z$-scores in Table \ref{epilepsytable}. Nonconjugate variational message passing compares well with the simulation-based full-data approach in terms of $z$-scores and is faster than both simulation-based approaches by an order of magnitude.
\vspace{-5mm}
\begin{table}[H]
\centering
\begin{footnotesize}
\begin{tabular}{*{3}{c}}
\multicolumn{3}{c}{Model I} \\
\hline Patient & $p_{i,\text{con}}^\text{CV}$ & $p_{i,\text{con}}^\text{NCVMP}$ \\ \hline
10 & 0.047 & 0.056 \\
25 & 0.048 & 0.062 \\
35 & 0.038 & 0.044 \\
56 & 0.023 & 0.028 \\
58 & 0.002 & 0.006 \\
\hline
\end{tabular} \hspace{3mm}
\begin{tabular}{*{3}{c}}
\multicolumn{3}{c}{Model II} \\
\hline Patient & $p_{i,\text{con}}^\text{CV}$ & $p_{i,\text{con}}^\text{NCVMP}$ \\ \hline
10 & 0.001 & 0.005 \\
25 & 0.024 & 0.049 \\
56 & 0.038 & 0.051 \\
\hline
\end{tabular}
\end{footnotesize}
\caption{\label{epilepsytable2} Epilepsy data. Conflict $p$-values for outliers in models I and II from cross-validatory approach and nonconjugate variational message passing using partially noncentered parametrization.}
\end{table}

At the 0.05 level, outliers identified by the cross-validatory approach are patients 10, 25, 35, 56 and 58 for Model I and patients 10, 25 and 56 for Model II. Table \ref{epilepsytable2} shows the cross-validatory conflict $p$-values for these patients. The corresponding conflict $p$-values computed using nonconjugate variational message passing with a partially noncentered parametrization are shown for comparison. While $p$-values from the two approaches are close, some of the outliers identified by the cross-validatory approach are not detected using nonconjugate variational message passing. One way to resolve this issue is to flag all patients with conflict $p$-values $<0.1$ say as possible outliers and recompute conflict $p$-values for this smaller group using cross-validatory approach. In this way, nonconjugate variational message passing can be regarded as a screening tool which will be very useful for large data sets.

\subsection{Polypharmacy data} \label{poly}
The polypharmacy data set \citep{Hosmer2013} contains data on 500 subjects studied over a period of seven years (available at \url{http://www.umass.edu/statdata/statdata/stat-logistic.html}). The outcome of interest is whether the subject is taking drugs from 3 or more different groups. The number of outpatient mental health visits (MHV) and inpatient mental health visits made by each subject were recorded each year. We consider the dummy variables MHV\textunderscore 1=1 if $1 \leq \text{MHV} \leq 5$ and 0 otherwise, MHV\textunderscore 2=1 if if $6 \leq \text{MHV} \leq 14$ and MHV\textunderscore 3=1 if $\text{MHV} \geq 15$ and 0 otherwise. Let INPTMHV = 0 if there were no inpatient mental health visits and 1 otherwise. Other covariates include Age, $\text{Gender}=1$ if male and 0 if female and $\text{Race}=0$ if subject is White and 1 otherwise. Following \cite{Hosmer2013}, we consider a logistic random intercept model of the form
\begin{multline}\label{polymodel}
\text{logit}(\mu_{ij})=\beta_0+\beta_1\text{Gender}_i+\beta_2 \text{Race}_i + \beta_3 \text{Age}_{ij}+ \beta_4 \text{MHV\textunderscore 1}_{ij}\\ + \beta_5 \text{MHV\textunderscore 2}_{ij} + \beta_6 \text{MHV\textunderscore 3}_{ij} + \beta_7 \text{INPTMHV}_{ij} + u_i,
\end{multline}
where $u_i \sim N(0,\sigma^2)$ for $i=1,\dots, 500$, $j=1,\dots,7$. 

\begin{figure}[H]
\centering
\includegraphics[width=0.96\textwidth]{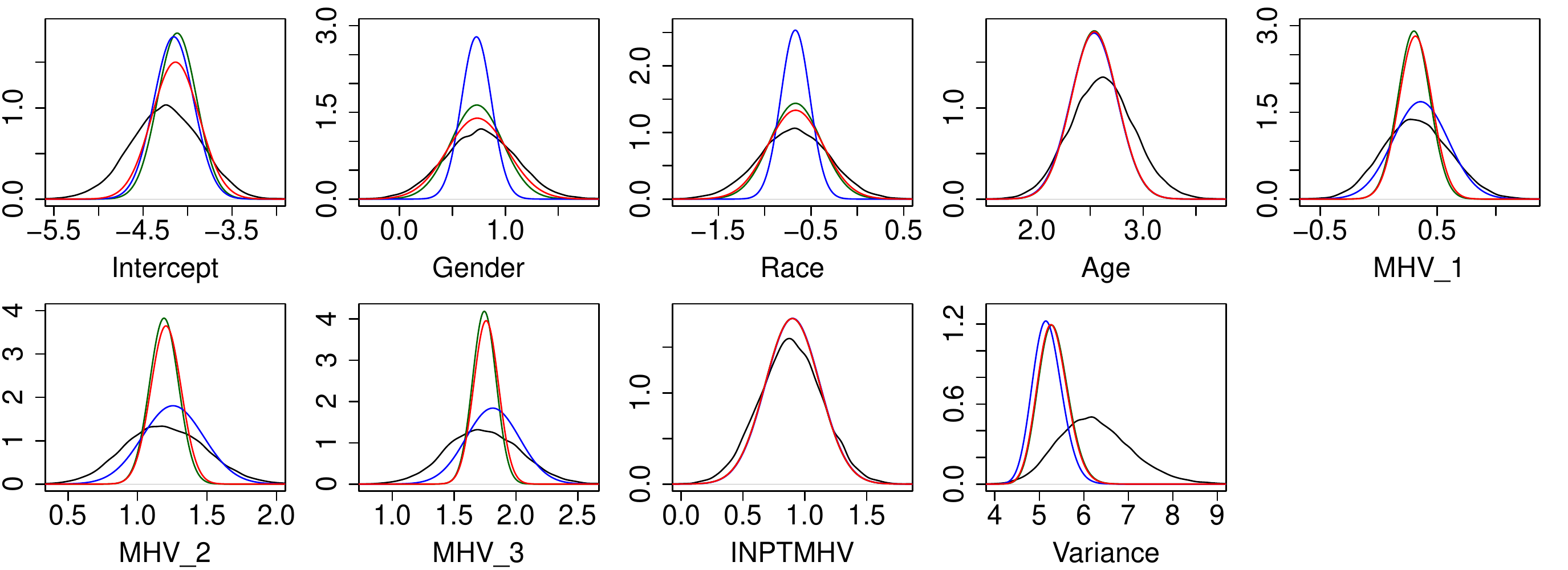}
\caption{\label{polyplot} Polypharmacy data. Marginal posteriors estimated by MCMC (black) and Algorithm \ref{Alg 1} using the centered (green), noncentered (blue) and partially noncentered (red) parametrizations.}
\end{figure}

This model was fitted using Algorithm \ref{Alg 1} and MCMC. Variational lower bounds and CPU times for model fitting are shown in Table \ref{polytable}. For MCMC, two chains were run simultaneously to assess convergence, each with 11,000 iterations, and the first 1000 iterations were discarded in each chain as burn-in. Algorithm 1 is of an order of magnitude faster than MCMC. Figure \ref{polyplot} shows the marginal posterior distributions of parameters estimated using MCMC and Algorithm 1. The partially noncentered parametrization attained the highest lower bound and took much less time to converge than the noncentered parametrization. Posterior approximations for $\beta_0$, $\beta_1$ and $\beta_2$ from partial noncentering were better than that of centering and noncentering. While posterior variance of $\beta_4$, $\beta_5$ and $\beta_6$ were underestimated by partial noncentering, the estimated posterior means were close to that of MCMC. As this data set is relatively small, using Algorithm \ref{Alg 2} in the initial stage of optimization did not lead to significant reductions in computation times.

\begin{table}[H]
\centering {\footnotesize \begin{tabular}{@{}lcccc@{}}  \hline
& noncentered  & centered  & partially noncentered & MCMC  \\  \hline
Lower bound ($\mathcal{L}$) & -1414.9 & -1414.4 & -1414.0 & -- \\
Time (model fitting) & 109.0  & 38.8 & 65.0 & 4320    \\  \hline
\end{tabular}
\caption{\label{polytable} Polypharmacy data. Variational lower bounds (first row) and CPU times (s) for model fitting (second row), for Algorithm \ref{Alg 1} (different parametrizations) and MCMC.}}
\end{table}
\vspace{-5mm}

To illustrate the improvements in efficiency that can be obtained from stochastic nonconjugate variational message passing, we simulated a larger data set comprising of $n=500 \times 20 = 10,000$ subjects from the model fitted by Algorithm \ref{Alg 1} (using the partially noncentered parametrization). The design matrices for each cluster were replicated 20 times and responses were generated from the model in \eqref{polymodel}, using as parameters variational posterior means from the fitted model. For this simulated data, Algorithm \ref{Alg 1} using the partially noncentered parametrization took 656.6 s to converge. 

\begin{figure}[H]
\centering
\includegraphics[width=0.95\textwidth]{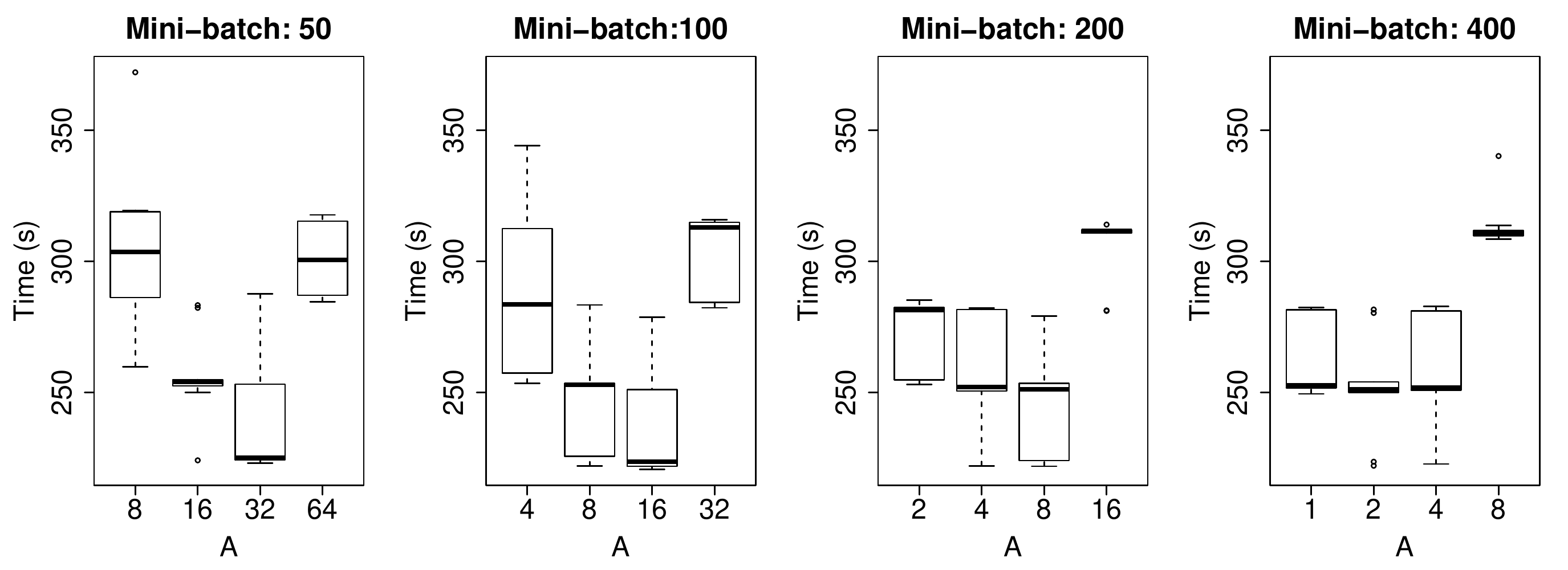}
\caption{\label{polyplot2} Polypharmacy simulated data. Computation times (s) for mini-batch sizes 50, 100, 200 and 400 (from left to right) corresponding to different stability constants displayed in boxplots.}
\end{figure}

For Algorithm \ref{Alg 2}, we considered mini-batch sizes $|B| \in \{50, 100, 200, 400\}$ (which correspond to 0.05\%, 1\%, 2\% and 4\% of $n=10,000$) and stability constants $A \in \{1,2,4,8,16, 32, 64\}$. Larger stability constants were used for smaller mini-batch sizes. For each mini-batch size and stability constant $A$, we performed ten runs of Algorithm \ref{Alg 2} switching to Algorithm \ref{Alg 1} when the relative increment in the lower bound after a sweep is less than $10^{-3}$. Computation times for the four mini-batch sizes corresponding to different stability constants are displayed in boxplots in Figure \ref{polyplot2}. The shortest average time to convergence for the different mini-batch sizes are given in Table \ref{polytable2} together with the corresponding stability constant $A$. From Figure \ref{polyplot2}, computation times were reduced by a factor of close to 2 or more across different mini-batch sizes and stability constants considered. Table \ref{polytable2} showed that larger stability constants $A$ are preferred for smaller mini-batch sizes. The shortest average time to convergence of 236.7 s was achieved by mini-batches of size 100 with $A=16$. This represents a reduction in computation time from Algorithm \ref{Alg 1} by a factor of 2.8.

\vspace{-3mm}
\begin{table}[H]
\centering 
 {\footnotesize\begin{tabular}{*{5}{c}}
\hline
$|B|$ & 50 & 100 & 200 & 400      \\  \hline  
$A$  & 32 & 16 & 8  & 2   \\ 
time  & 239.6 & 236.7 & 246.0 & 251.9  \\ \hline
\end{tabular}
\caption{\label{polytable2} Polypharmacy simulated data. Shortest average time to convergence (s) for different mini-batch sizes  together with corresponding stability constant $A$.}}
\end{table}
\vspace{-3mm}
\begin{figure}[H]
\centering
\includegraphics[width=0.57\textwidth]{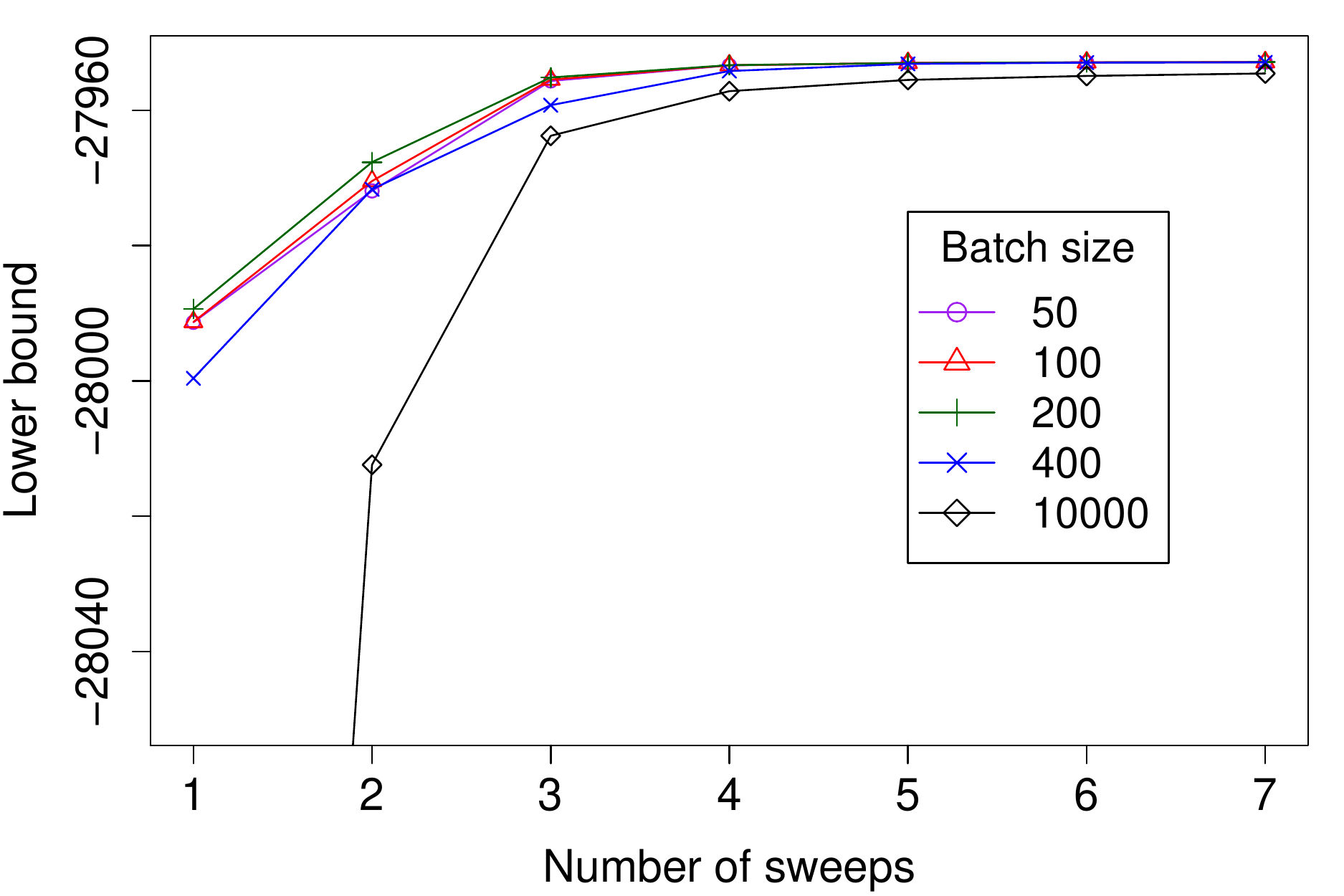}
\caption{\label{polyplot3}Polypharmacy simulated data. Plot of average lower bound against number of sweeps for different batch sizes, with stability constants $A$ given in Table \ref{polytable2}.}
\end{figure}
Figure \ref{polyplot3} tracks the average lower bound attained at the end of each sweep for different mini-batch sizes, with stability constants $A$ given in Table \ref{polytable2}. Only the first seven sweeps are shown. Figure \ref{polyplot3} shows that with appropriate step sizes, stochastic nonconjugate variational message passing is able to make much bigger gains than the standard version, particularly in the first few sweeps. Thus, for moderate-sized data sets, gains in computation times can be obtained by using Algorithm \ref{Alg 2} in the initial stage of optimization.

\subsection{Skin cancer prevention study} \label{skincancer}
In a clinical trial to test the effectiveness of beta-carotene in preventing non-melanoma skin cancer \citep{Greenberg1989}, 1805 high risk patients were randomly assigned to receive either a placebo or 50 mg of beta-carotene per day for five years. The response $y_{ij}$ is a count of the number of new skin cancers in year $j$ for the $i$th subject. Covariate information for the $i$th subject include $\text{Age}_i$, the age in years at the beginning of the study, $\text{Gender}_i=1$ if male and 0 if female, $\text{Skin}_i=1$ if skin has burns and 0 otherwise, $\text{Exposure}_i$, a count of the number of previous skin cancers, and $\text{Year}_{ij}$, the year of follow-up. The treatment effect has been shown to be insignificant in previous analyses. We consider $n=1683$ subjects with complete covariate information (data set available at \url{http://www.biostat.harvard.edu/~fitzmaur/ala2e/}). Following \citet{Donohue2011}, we consider the random intercept and slope model 
\begin{equation}\label{skinmodel}
\log(\mu_{ij})=\beta_0 +\beta_1 \text{Year}_{ij} +\beta_2 \text{Age}_i+\beta_3 \text{Gender}_i  +\beta_4 \text{Skin}_i+\beta_5 \text{Exposure}_i +u_{1i}+u_{2i}\text{Year}_{ij},
\end{equation}
where $\left[\begin{smallmatrix} u_{1i}\\u_{2i} \end{smallmatrix}\right]\sim N\left(0,\left[\begin{smallmatrix} \sigma_{11}^2 & \sigma_{12}\\\sigma_{21} & \sigma_{22}^2 \end{smallmatrix}\right]\right)$ for $i=1,\dots, 1683$, $1 \leq j \leq 5$. The covariates Year, Age and Skin were standardized to have mean 0 and variance 1.

\begin{figure}[H]
\centering
\includegraphics[width=0.85\textwidth]{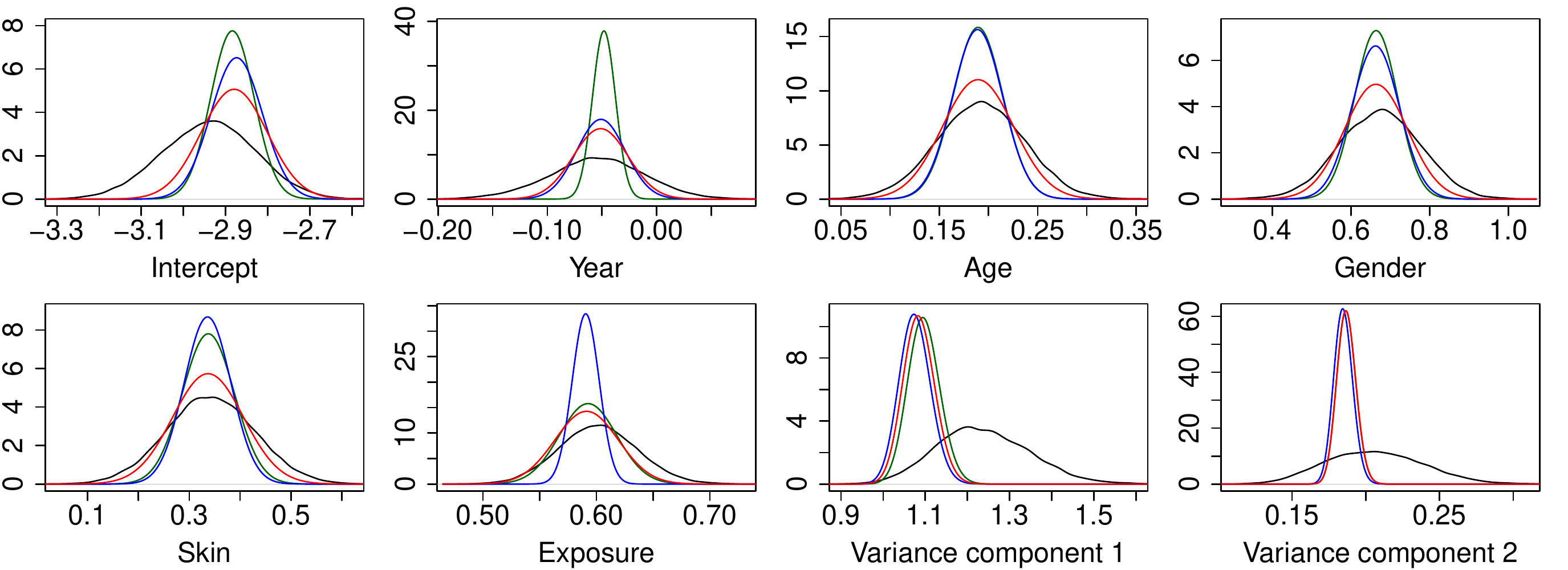}
\caption{\label{skinplot} Skin cancer data. Marginal posteriors estimated by MCMC (black) and Algorithm \ref{Alg 1} using the centered (green), noncentered (blue) and partially noncentered (red) parametrizations.}
\end{figure}

Fitting this model using Algorithm \ref{Alg 1} and MCMC, the estimated marginal posterior distributions of model parameters are shown in Figure \ref{skinplot} and computation times and variational lower bounds are given in Table \ref{skintable}. For MCMC, two chains were run simultaneously to assess convergence, each with 11,000 iterations, and the first 1000 iterations were discarded in each chain as burn-in. Partial noncentering performed very well as compared to centering and noncentering, producing posterior approximations that were closest to that of MCMC and converging in the shortest time.
\begin{table}[H]
\centering {\footnotesize \begin{tabular}{@{}lccKc@{}}  \hline
& noncentered  & centered  & partially noncentered & MCMC  \\  \hline
Lower bound ($\mathcal{L}$) & -4054.1 & -4054.1 & -4051.7 & -- \\
Time (model fitting) & 46.6 & 42.6 & 42.0 & 11113 \\  \hline
\end{tabular}
\caption{\label{skintable} Skin cancer data. Variational lower bounds (first row) and CPU times (s) for model fitting (second row), for Algorithm \ref{Alg 1} (different parametrizations) and MCMC.}}
\end{table}

To investigate the performance of stochastic nonconjugate variational message passing, we simulated a much larger data set (comprising of $n=1683 \times 15 = 25245$ subjects) from the model fitted by Algorithm \ref{Alg 1} (using the partially noncentered parametrization). The design matrices for each cluster were replicated by 15 times and responses were generated from model \eqref{skinmodel} using as parameters variational posterior means from the fitted model. For large data sets, penalized quasi-likelihood may not be feasible for use as initialization as they converge too slowly (e.g. penalized quasi-likelihhood took more than 9 mins to converge for this simulated data set). Using the fit from GLM as initialization, Algorithm \ref{Alg 1} (using the partially noncentered parametrization) took 1230.9 s to converge. 

\begin{figure}[H]
\centering
\includegraphics[width=0.9\textwidth]{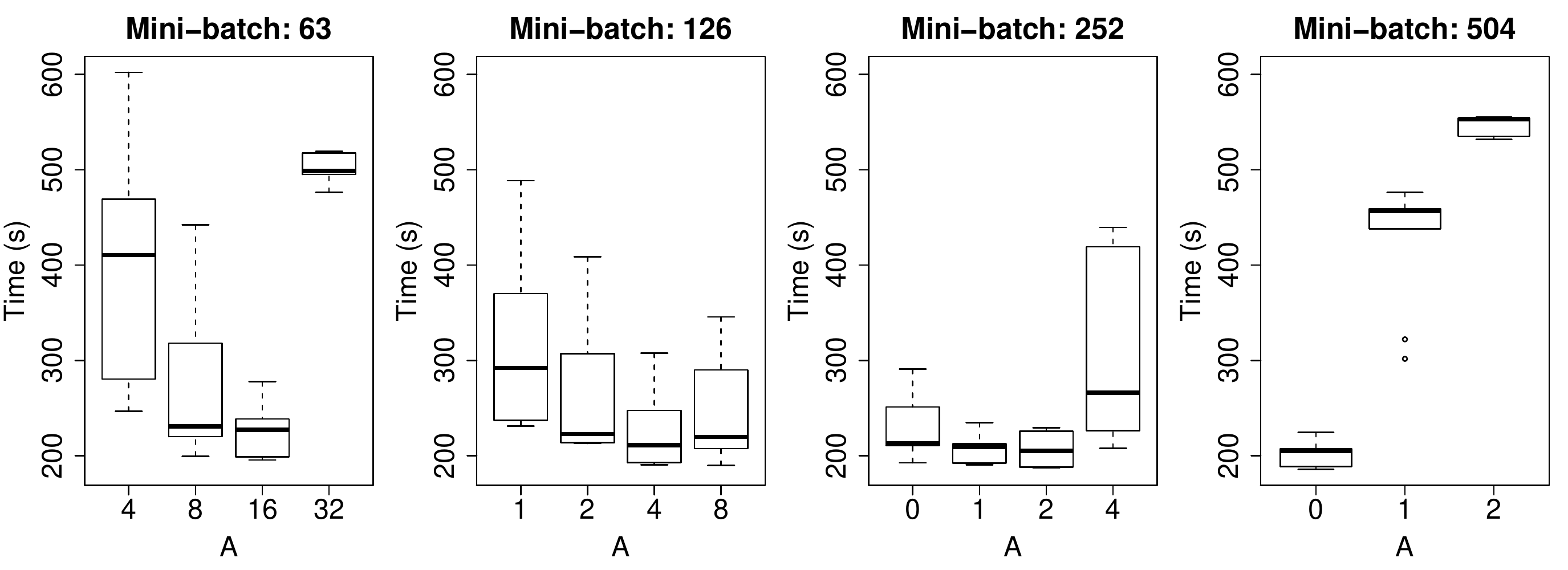}
\caption{\label{skinplot2} Skin cancer simulated data. Computation times (s) for mini-batch sizes 63, 126, 252 and 504 (from left to right) corresponding to different stability constants displayed in boxplots.}
\end{figure}
\vspace{-3mm}
\begin{table}[H]
\centering 
 {\footnotesize\begin{tabular}{*{5}{c}}
\hline
$|B|$  & 63 & 126 & 252 & 504     \\  \hline  
$A$   & 8 & 4 & 2 & 0  \\ 
time  & 266.3  & 224.4 & 205.3 & 200.8 \\ \hline
\end{tabular}
\caption{\label{skintable2} Skin cancer simulated data. Shortest average time to convergence (s) for different mini-batch sizes  together with corresponding stability constant $A$.}}
\end{table}
We consider mini-batch sizes $|B| \in \{63,126,252,504\}$ (corresponding to 0.025\%, 0.05\%, 1\%, and 2\% of $n=25245$) and stability constants $A \in \{0,1,2,4,8,16,32\}$. Larger stability constants were used for smaller mini-batch sizes. For each mini-batch size and stability constant, we performed ten runs of Algorithm \ref{Alg 2}, switching to Algorithm \ref{Alg 1} when the relative increment in the lower bound after a sweep is less than $10^{-3}$. Computation times for the four mini-batch sizes corresponding to different stability constants are displayed in boxplots in Figure \ref{skinplot2}. The shortest average time to convergence for different mini-batch sizes are given in Table \ref{skintable2} together with the corresponding stability constant $A$. From Figure \ref{skinplot2}, computation times were reduced by a factor of 2 or more across the different mini-batch sizes and stability constants considered. As in the previous example, Table \ref{skintable2} showed that larger stability constants $A$ are preferred for smaller mini-batch sizes. The shortest average time to convergence of 200.8 s was achieved by mini-batches of size 504 with $A=0$. This represents a reduction in computation time from Algorithm \ref{Alg 1} by a factor of 6. Similar results can be achieved by smaller mini-batch sizes with appropriately chosen step sizes.

\begin{figure}[H]
\centering
\includegraphics[width=0.65\textwidth]{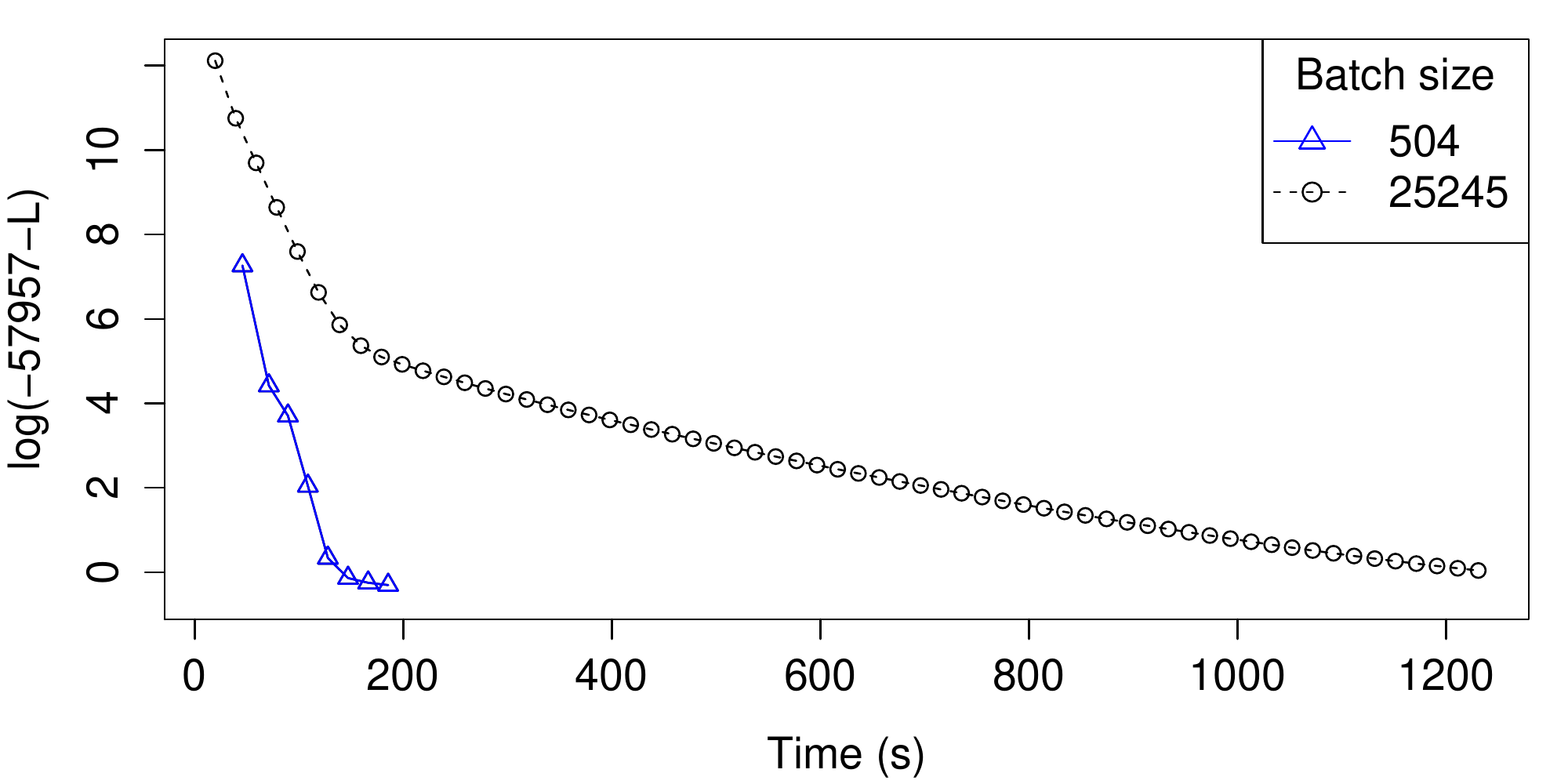}
\caption{\label{skinlb}Skin cancer simulated data. Plot of $\log(-57957-\mathcal{L})$ against time for the mini-batch of size 504 ($A=0$) fitted using Algorithm \ref{Alg 2} in the initial stage followed by Algorithm \ref{Alg 1} and the whole data set fitted using Algorithm \ref{Alg 1}.}
\end{figure}
Figure \ref{skinlb} compares the rate of convergence of standard and stochastic nonconjugate variational message passing for one of the runs where $|B|=504$ and $A=0$. The variational lower bound $\mathcal{L}$ is $-57958$ at convergence and $\log(-57957-\mathcal{L})$ is plotted against time. Stochastic nonconjugate variational message passing took just 8 sweeps to converge in 208.0 s while standard nonconjugate variational message passing took 62 sweeps and converged in 1230.9 seconds. This represents a reduction in computation time by a factor of close to 6.

\section{Conclusion}
In this paper, we have extended stochastic variational inference to nonconjugate models and derived a stochastic nonconjugate variational message passing algorithm that is scalable to large data sets. The data sets that we have considered in this paper were only of moderate size. Nevertheless, we show that computation times can be reduced by applying stochastic nonconjugate variational message passing in the initial stage of optimization. The stochastic version seems computationally preferable once the number of clusters is of the order of ten thousand and above. We imagine the gain to be bigger for larger data sets and more work remains to be done in that aspect. Experimentation with various settings of stability constants $A$ suggest that larger $A$ is preferred for smaller mini-batch sizes. To avoid hand-tuning of step sizes, it will be useful to develop adaptive step sizes for stochastic nonconjugate variational message passing and we are currently working on extending the work of \cite{Ranganath2013} to nonconjuagte models. We have also shown that conflict diagnostics for identifying divergent units can be obtained as a by-product of nonconjugate variational message passing. Our diagnostics approximate the approach of \citet{Marshall2007} and experiments suggest relatively good agreement between the two methods. For large data sets, computation of conflict $p$-values using simulation-based approaches is very computationally intensive and nonconjugate variational message passing is attractive as an alternative for obtaining prior-likelihood diagnostics or for use as an initial screening tool.

{\footnotesize
\section*{Acknowledgements}
Linda Tan was partially supported as part of the Singapore-Delft Water Alliance's tropical reservoir research programme. We thank Matt Wand for making available to us his preliminary work on fully simplified multivariate normal non-conjugate variational message passing updates. 
}

\appendix

\section{Simplified updates and natural gradients for conjugate \\factors in nonconjugate variational message passing}

Let $N(\theta_i)$ denote the neighbourhood of $\theta_i$ in the factor graph of $p(y,\theta)$ \citep[see][]{Tan2013}.
Suppose $p(y,\theta)=\prod_a f_a(y,\theta)$ and each factor $f_a$ in $N(\theta_i)$ is conjugate to $q_i(\theta_i|\lambda_i)$, say \begin{equation*}
f_a(y,\theta)= \exp\left\{g_a(y,\theta_{-i})^T t_i(\theta_i)-h_a(y,\theta_{-i})\right\},
\end{equation*}
where $\theta_{-i}=(\theta_1,\dots,\theta_{i-1},\theta_{i+1},\dots,\theta_m)$. Then 
\begin{equation*}
{\nabla}_{\lambda_i} \mathcal{L} =\mathcal{V}_i(\lambda_i) \left[ \sum\nolimits_{a \in N(\theta_i)} E_q\{g_a(y,\theta_{-i})\}-\lambda_i \right]
\end{equation*}
and the nonconjugate variational message passing update in \eqref{NCVMP_update} reduces to 
\begin{equation}\label{VMP_update}
\lambda_i \leftarrow \sum\nolimits_{a \in N(\theta_i)} E_q\{g_a(y,\theta_{-i})\}.
\end{equation}
Note that $E_q\{g_a(y,\theta_{-i})\}$ does not depend on $\lambda_i$. The natural gradient in \eqref{nat_grad} can also be simplified as
\begin{equation} \label{VMP_nat_grad}
\widetilde{\nabla}_{\lambda_i}\mathcal{L} =\sum_{a \in N(\theta_i)}E_q\{g_a(y,\theta_{-i})\}-\lambda_i.
\end{equation}
 
\section{Definition of notation and derivation of updates in \\Algorithm 2}
For Poisson responses, 
\begin{equation*}
g_i=E_i\odot \exp \left\{V_i \mu_{q(\beta)} + Z_i \mu_{q(\tilde{\alpha}_i)}+\tfrac{1}{2}\text{diag}(V_i \Sigma_{q(\beta)} {V_i}^T+Z_i\Sigma_{q(\tilde{\alpha}_i)}{Z_i}^T)\right\} \; \text{and} \; F_i=\text{diag}(g_i)
\end{equation*}
for $i=1,\dots,n$. For Bernoulli responses,
\begin{equation*}
g_i=B^{(1)}(\mu_i^q, \sigma_i^q) \; \text{and}\; F_i=\text{diag}(B^{(2)}(\mu_i^q, \sigma_i^q)) 
\end{equation*}
for $i=1,\dots,n$, where $\mu_i^q=V_{i}\mu_{q(\beta)} +Z_i \mu_{q(\tilde{\alpha}_i)} $ and $\sigma_i^q=\sqrt{\text{diag}(V_{i} \Sigma_{q(\beta)} V_{i}^T +Z_i \Sigma_{q(\tilde{\alpha}_i)} Z_i^T)}$. We have 
\begin{equation*}
B^{(r)}(\mu,\sigma)=\int_{-\infty}^{\infty} b^{(r)}(\sigma x+\mu)\frac{1}{\sqrt{2\pi}}\exp(-x^2)\;dx,
\end{equation*} 
where $b(x)=\log\{1+\exp(x)\}$ and $b^{(r)}(x)$ denotes the $r$th derivative of $b(\cdot)$ with respect to $x$. If $\mu$ and $\sigma$ are vectors, say $\mu=\left[\begin{smallmatrix}1\\2\\3 \end{smallmatrix}\right]$ and $\sigma=\left[\begin{smallmatrix}4\\5\\6 \end{smallmatrix}\right]$, then
$B^{(r)} (\mu,\sigma)=\left[\begin{smallmatrix} B^{(r)}(1,4)\\B^{(r)}(2,5)\\B^{(r)}(3,6)\end{smallmatrix}\right]$.  The terms, $B^{(r)} (\mu,\sigma)$, $r=0,1,2$ may be evaluated efficiently using adaptive Gauss-Hermite quadrature \citep{Liu1994}. More details can be found in \cite{Tan2013}. 

The updates in step 2 of Algorithm \ref{Alg 2} are taken directly from the nonconjugate variational message passing algorithm for GLMMs in \cite{Tan2013}. To derive the updates in step 3, let us first introduce the following notation for specification of the natural parameter vectors $\lambda_\beta$ and $\lambda_D$. For a $d \times d$ square matrix $A$, let $\text{vec}(A)$ denote the $d^2 \times 1$ vector obtained by stacking the columns of $A$ under each other, from left to right in order and $\text{vech}(A)$ denotes the $\tfrac{1}{2}d(d+1) \times 1$ vector obtained from $\text{vec}(A)$ by eliminating all supradiagonal elements of $A$. The matrix $D_d$ is a unique $d^2 \times \tfrac{1}{2}d(d+1)$ matrix that transforms $\text{vech}(A)$ into $\text{vec}(A)$ if $A$ is symmetric, that is, $D_d\text{vech}(A)=\text{vec}(A)$. See \cite{Magnus1988} for more details. We have
\begin{equation*}
\lambda_\beta=\begin{bmatrix} -\frac{1}{2}D_p^T \text{vec}(\Sigma_{q(\beta)}^{-1}) \\ \Sigma_{q(\beta)}^{-1}\mu_{q(\beta)} \end{bmatrix}\;\;\text{and}\;\;\lambda_D=\begin{bmatrix}  -\frac{1}{2}\text{vec}(S_{q(D)}) \\  -\frac{1}{2} (\nu_{q(D)}+r+1) \end{bmatrix}.
\end{equation*}
From \eqref{previousiterate}, 
\begin{equation}\label{betaupdate}
\begin{bmatrix} -\frac{1}{2}D_p^T \text{vec}({\Sigma_{q(\beta)}^{(t+1)}}^{-1}) \\ {\Sigma_{q(\beta)}^{(t+1)}}^{-1}\mu_{q(\beta)}^{(t+1)} \end{bmatrix} = (1-a_t) \begin{bmatrix} -\frac{1}{2}D_p^T \text{vec}({\Sigma_{q(\beta)}^{(t)}}^{-1}) \\ 
{\Sigma_{q(\beta)}^{(t)}}^{-1}\mu_{q(\beta)}^{(t)}\end{bmatrix} + a_t \begin{bmatrix} -\frac{1}{2}D_p^T \text{vec}(\hat{\Sigma}_{q(\beta)}^{-1}) \\ \hat{\Sigma}_{q(\beta)}^{-1} \hat{\mu}_{q(\beta)} \end{bmatrix},
\end{equation}
where
\begin{align*}
\hat{\Sigma}_{q(\beta)} &= \left[ \Sigma_\beta^{-1} + \frac{n}{|B|}\sum_{i\in B} \big\{ \nu_{q(D)} \tilde{W}_i^TS_{q(D)}^{-1}\tilde{W}_i + V_i^T F_i V_i  \big\} \right]^{-1} \\
\hat{\mu}_{q(\beta)} &= \mu_{q(\beta)}^{(t)} + \hat{\Sigma}_{q(\beta)}\left[  \frac{n}{|B|}\sum_{i\in B} \big\{ \nu_{q(D)} {\tilde{W}_i}^TS_{q(D)}^{-1}(\mu_{q(\tilde{\alpha}_i)}-\tilde{W}_i\mu_{q(\beta)}^{(t)} )+ V_i^T(y_i-G_i) \big\}-\Sigma_\beta^{-1}\mu_{q(\beta)}^{(t)} \right].
\end{align*}
Expressions for $\hat{\Sigma}_{q(\beta)}$ and $\hat{\mu}_{q(\beta)}$ can be deduced from Algorithm 3 of \cite{Tan2013}. The first line in \eqref{betaupdate} gives 
\begin{equation*}
{\Sigma_{q(\beta)}^{(t+1)}}=\left\{   (1-a_t){\Sigma_{q(\beta)}^{(t)}}^{-1}+a_t \hat{\Sigma}_{q(\beta)}^{-1}\right\}^{-1},
\end{equation*}
which is the update for $\Sigma_{q(\beta)}$ in Algorithm 2.
The second line in \eqref{betaupdate} gives 
\begin{align*}
\mu_{q(\beta)}^{(t+1)} &= {\Sigma_{q(\beta)}^{(t+1)}}\left\{  (1-a_t)  {\Sigma_{q(\beta)}^{(t)}}^{-1}\mu_{q(\beta)}^{(t)} +a_t \hat{\Sigma}_{q(\beta)}^{-1} \hat{\mu}_{q(\beta)}  \right\} \\
&= {\Sigma_{q(\beta)}^{(t+1)}}\left\{ \left({\Sigma_{q(\beta)}^{(t+1)}}^{-1}-a_t \hat{\Sigma}_{q(\beta)}^{-1} \right) \mu_{q(\beta)}^{(t)} +a_t \hat{\Sigma}_{q(\beta)}^{-1} \hat{\mu}_{q(\beta)}  \right\} \\
&= \mu_{q(\beta)}^{(t)} + a_t  {\Sigma_{q(\beta)}^{(t+1)}} \hat{\Sigma}_{q(\beta)}^{-1} \left( \hat{\mu}_{q(\beta)} - \mu_{q(\beta)}^{(t)} \right)  \\
&= \mu_{q(\beta)}^{(t)} + a_t  {\Sigma_{q(\beta)}^{(t+1)}} \left[  \frac{n}{|B|}\sum_{i\in B} \big\{ \nu_{q(D)} {\tilde{W}_i}^TS_{q(D)}^{-1}(\mu_{q(\tilde{\alpha}_i)}-\tilde{W}_i\mu_{q(\beta)}^{(t)} )+ V_i^T(y_i-G_i) \big\}-\Sigma_\beta^{-1}\mu_{q(\beta)}^{(t)} \right],
\end{align*}
which is the update for $\mu_{q(\beta)}$ in Algorithm 2. Similarly, from \eqref{previousiterate}, we have 
\begin{equation}\label{Dupdate}
\begin{bmatrix}  -\frac{1}{2}\text{vec}(S_{q(D)}^{(t+1)}) \\  -\frac{1}{2} (\nu_{q(D)}^{(t+1)}+r+1) \end{bmatrix}= (1-a_t)\begin{bmatrix}  -\frac{1}{2}\text{vec}(S_{q(D)}^{(t)}) \\  -\frac{1}{2} (\nu_{q(D)}^{(t)}+r+1) \end{bmatrix}+ a_t\begin{bmatrix}  -\frac{1}{2}\text{vec}(\hat{S}_{q(D)}) \\  -\frac{1}{2} (\hat{\nu}_{q(D)}+r+1) \end{bmatrix},
\end{equation}
where $\hat{S}_{q(D)}$ and $\hat{\nu}_{q(D)}$ can be deduced from \cite{Tan2013} as $\hat{\nu}_{q(D)} = \nu+n$ and $\hat{S}_{q(D)} =S+ \frac{n}{|B|}\sum_{i\in B} \big\{ (\mu_{q(\tilde{\alpha}_i)}-\tilde{W}_i \mu_{q(\beta)}) (\mu_{q(\tilde{\alpha}_i)}-\tilde{W}_i \mu_{q(\beta)})^T + \Sigma_{q(\tilde{\alpha}_i)}+\tilde{W}_i\Sigma_{q(\beta)}\tilde{W}_i^T \big\} $. The updates for $S_{q(D)}$ and $\nu_{q(D)}$ in Algorithm 2 can be obtained by simplifying \eqref{Dupdate}.

\section{Generating likelihood replicates}
In the centered parametrization,
\begin{equation*}
\eta_i = Z_i \alpha_i + X_{gi} \beta_g \;\;\text{where}\;\; \alpha_i = C_i \beta_c + u_i \sim N(C_i \beta_c, D)
\end{equation*}
for $i=1, \dots,n$. To generate likelihood replicates $\alpha_i^{\text{lik}}$ from $p(\alpha_i|y_i)$ in the cross-validatory approach, we consider Jeffreys's prior for the centered random effects $\alpha_i$. Jeffreys's prior is defined as $p(\alpha_i) \propto \sqrt{|I(\alpha_i)|}$, where $I(\alpha_i)$ is the Fisher information matrix of $\alpha_i$. For Poisson and logistic GLMMs, it can be shown that $p(\alpha_i) \propto |Z_i^T Q_i Z_i|^{\frac{1}{2}}$, where $Q_i$ is a $n_i \times n_i$ diagonal matrix \cite[see, e.g.][]{Ibrahim1991}. Definitions of $Q_i$ are given in Section \ref{Sec_PNCP}. In general, we will need to consider $\beta_g$ as a nuisance parameter. Following the discussion in Section \ref{CV}, we generate a $\beta_g$ from $p(\beta_g|y_{-i})$ and simulate $\alpha_i^{\text{lik}}$ from $p(y_i|\alpha_i,\beta^g)p(\alpha_i)$ where $p(\alpha_i)$ is Jeffreys's prior. For Poisson GLMMs, 
\begin{equation*}
p(y_i|\alpha_i,\beta^g)p(\alpha_i) \propto \exp\{ y_i^T(\log E_i + Z_i \alpha_i + X_{gi} \beta_g)-E_i^T\exp(Z_i \alpha_i + X_{gi} \beta_g) \} |Z_i^T Q_i Z_i|^{\frac{1}{2}}.
\end{equation*}
For logistic GLMMs,
\begin{equation*}
p(y_i|\alpha_i,\beta^g)p(\alpha_i) \propto \exp[ y_i^T(Z_i \alpha_i + X_{gi} \beta_g)-1_{n_i}^T\log\{1_{n_i}+\exp(Z_i \alpha_i + X_{gi} \beta_g) \} ] |Z_i^T Q_i Z_i|^{\frac{1}{2}}.
\end{equation*}

\section{Additional Example: Madras schizophrenia data} \label{schizo}

The Madras schizophrenia study \citep{Thara1994} contains records of the psychiatric symptoms of 86 patients in the first year after initial hospitalization. This data set has been analyzed by \cite{Diggle2002} and is available at \url{http://faculty.washington.edu/heagerty/Books/AnalysisLongitudinal/datasets.html}. The reponse $y_{ij}$ is 1 if the symptom ``thought disorder'' is present and 0 otherwise. We consider the covariates, age at onset of disease (Age = 1 if patient is at least 20 years old and 0 otherwise), sex of patient (Gender = 1 if female and 0 otherwise) and number of months since hospitalization when symptom was recorded ($t$). We consider the logistic random effects model:
\begin{equation*}
\text{logit}(\mu_{ij})=\beta_0 + \beta_1\text{Age}_i + \beta_2 \text{Gender}_i +\beta_3 t_{ij} + \beta_4 \text{Age}_i\times t_{ij}  + \beta_5 \text{Gender}_i \times t_{ij}  + u_i,
\end{equation*}
 where $u_i\sim N(0,\sigma^2)$ for $i=1,\dots,86$, $1\leq j \leq12$. We report both one-sided (upper-tail) and two-sided conflict $p$-values for this example. The upper-tail one-sided conflict $p$-values are useful for identifying patients with unusually large number of ``thought disorders" while the two-sided conflict $p$-values can be used to detect patients with either more or less than the expected number of ``thought disorders".

In the simulation-based approaches, $\beta_3$, $\beta_4$ and $ \beta_5$ have to be regarded as nuisance parameters under the centered parametrization (see Appendix C). For each model fitting via MCMC, two chains were run simultaneously to assess convergence, each with 26,000 iterations, and the first 1000 iterations were discarded in each chain as burn-in. Simulation-based conflict $p$-values were calculated based on the remaining 50,000 simulations. For the cross-validatory approach, model refitting took a total of 372 s $\times$ 86 (more than 8 hours) to complete in OpenBUGS. Simulation of prior and likelihood replicates of the centered random effects $\alpha_i$ was performed in {\ttfamily R}. Assuming Jeffreys's prior for $\alpha_i$, likelihood replicates were simulated using adaptive rejection metropolis sampling.

\begin{figure}[H]
\centering
\includegraphics[width=0.93\textwidth]{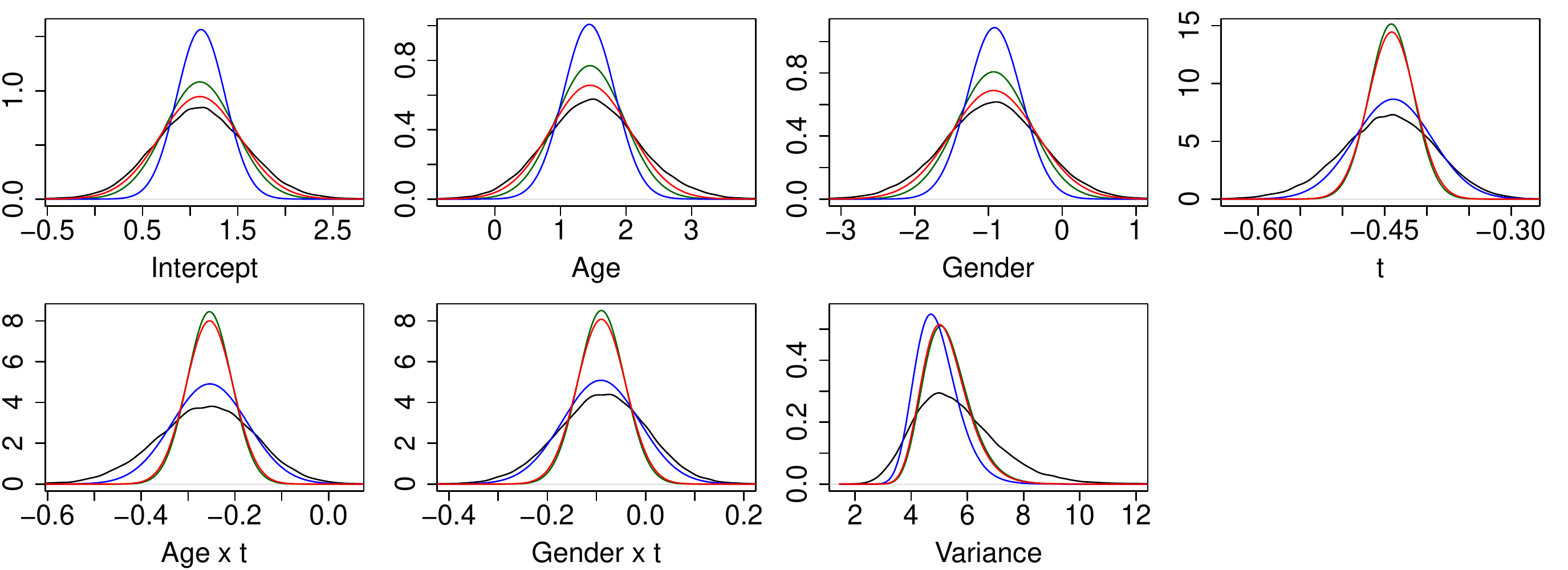}
\caption{\label{madrasplot} Madras data. Marginal posteriors estimated by MCMC (black) and Algorithm \ref{Alg 1} using the centered (green), noncentered (blue) and partially noncentered (red) parametrizations.}
\end{figure}
\vspace{-5mm}
\begin{table}[H]
\centering {\footnotesize \begin{tabular}{@{}lccKK@{}}  \hline
& noncentered  & centered  & partially noncentered & MCMC (full-data)  \\  \hline
Lower bound ($\mathcal{L}$) & -407.9 & -407.1 & -406.6 & -- \\
Time (model fitting) & 6.4  & 6.0 & 5.0 & 372    \\ 
Time (computing conflict $p$-values) & 0.1 & 0.1 & 0.1 & 16266.9  \\
Mean absolute difference in $z$-scores (one-sided) & 0.115 & 0.102 & 0.104 & 0.040 \\
Mean absolute difference in $z$-scores (two-sided) & 0.227 & 0.201 & 0.204 & 0.069 \\  \hline
\end{tabular}
\caption{\label{madrastable} Madras data. Variational lower bounds (first row), CPU times (s) for model fitting (second row) and calculating conflict $p$-values (third row) and mean absolute difference in $z$-scores (relative to cross-validatory approach) for one-sided (fourth row) and two-sided (fifth row) $p$-values, for Algorithm \ref{Alg 1} (different parametrizations) and MCMC (full-data). }}
\end{table}

Variational lower bounds and CPU times taken for model fitting and computing conflict $p$-values by Algorithm \ref{Alg 1} (different parametrizations) and MCMC (full-data approach) are given in Table \ref{madrastable}. Figure \ref{madrasplot} shows the marginal posteriors of parameters estimated using MCMC and Algorithm \ref{Alg 1}. The partially noncentered parametrization took the shortest time to converge and attained the highest lower bound. From Figure \ref{madrasplot}, partial noncentering produced better posterior approximations for $\beta_0$, $\beta_1$ and $\beta_2$ than both centering and noncentering. For $\beta_3$, $\beta_4$, $\beta_5$, partial centering performed better than centering but did not do as well as noncentering.

\begin{figure}[H]
\centering
\includegraphics[width=0.8\textwidth]{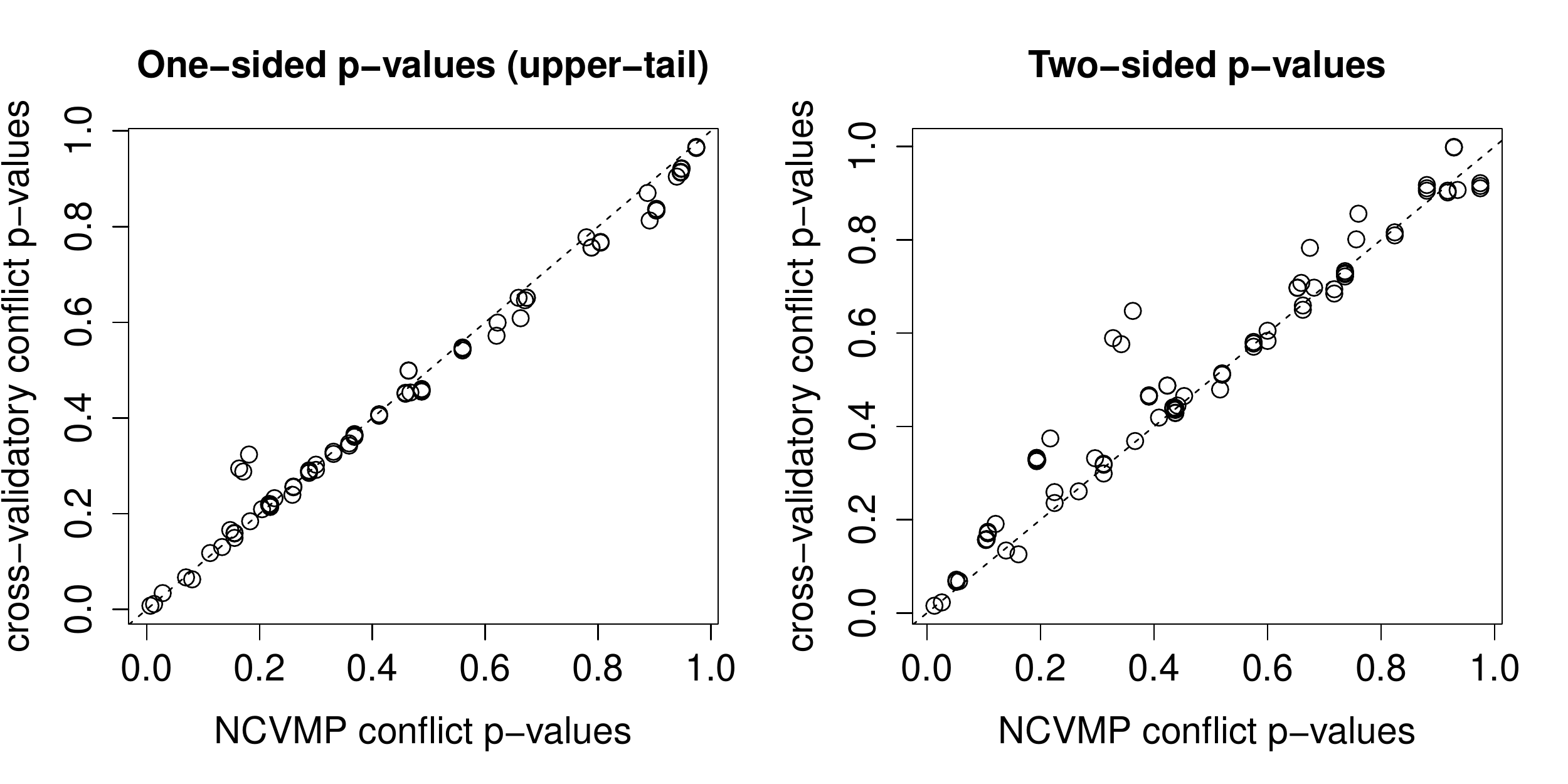}
\caption{ \label{madrascv} Madras data. Cross-validatory conflict $p$-values plotted against conflict $p$-values from nonconjugate variational message passing with a partially noncentered parametrization.}
\end{figure}

Cross-validatory conflict $p$-values are plotted against conflict $p$-values from nonconjugate variational message passing using the partially noncentered parametrization in Figure \ref{madrascv}. The left plot shows the upper-tail one-sided $p$-values while the right plot shows the two-sided $p$-values. The mean absolute difference in $z$-scores for nonconjugate variational message passing and the simulation-based full-data approach relative to the cross-validatory approach are given in Table \ref{madrastable}. Figure \ref{madrascv} shows that the agreement between the cross-validatory approach and nonconjugate variational message passing is better for the one-sided $p$-values than in the two-sided case. This is expected as any discrepancy between the two sets of $p$-values will be doubled in the two-sided case. However, we note that agreement at the extremes is still relatively good. For this example, the simulation-based full-data approach performed better in terms of $z$-scores than nonconjugate variational message passing. This is likely due to the fact that in this case, the variational posterior does not provide as good an approximation to the true posterior as in Examples 6.1 and 6.2. However, nonconjugate variational message passing remains useful as a screening tool as the computation time required to compute conflict $p$-values even in the simulation-based full-data approach is quite significant. Finally, outliers (at the 0.05 level) identified by the cross-validatory approach and nonconjugate variational message passing using the partially noncentered parametrization are identical in this example. Conflict $p$-values for these outliers are shown in Table \ref{madrastable2}.

\vspace{-5mm}
\begin{table}[H]
\centering \ra{1.08}
\begin{footnotesize}
\begin{tabular}{*{3}{c}}
\multicolumn{3}{c}{One-sided $p$-values (upper-tail)} \\
\hline Patient & $p_{i,\text{con}}^\text{CV}$ & $p_{i,\text{con}}^\text{NCVMP}$ \\ \hline
14 & 0.034 & 0.028 \\
27 & 0.011 & 0.013 \\
68 & 0.008 & 0.007 \\ \hline
\end{tabular} \hspace{3mm}
\begin{tabular}{*{3}{c}}
\multicolumn{3}{c}{Two-sided $p$-values} \\
\hline Patient & $p_{i,\text{con}}^\text{CV}$ & $p_{i,\text{con}}^\text{NCVMP}$ \\ \hline
25 & 0.023 & 0.026 \\
56 & 0.016 & 0.013 \\ \hline
\end{tabular}
\end{footnotesize}
\caption{\label{madrastable2} Madras data. Conflict $p$-values for outliers from cross-validatory approach and nonconjugate variational message passing using partially noncentered parametrization.}
\end{table}

\end{document}